\documentclass{aa}
\def\apgt{\ {\raise-.5ex\hbox{$\buildrel>\over\sim$}}\ }
\def\aplt{\ {\raise-.5ex\hbox{$\buildrel<\over\sim$}}\ }
\def\lteq{\ {\raise-.5ex\hbox{$\buildrel<\over-$}}\ }
\newcommand{\MSun}{\mbox{${\rm M}_\odot$}}
\newcommand{\MJup}{\mbox{${\rm M}_{\rm Jup}$}}
\newcommand{\address}[1]{\gdef\@address{#1}}%

\usepackage{multicol}
\usepackage{booktabs}
\usepackage{array}

\usepackage{graphicx}
\usepackage{listings}
\usepackage{pgfplots}
\pgfplotsset{compat=1.9}
\usepackage{pgf}
\usepackage{gensymb}
\usepackage{txfonts}

\newcommand{\solus}{\emph{s\={o}lus lapis}}
\newcommand{\soli}{\emph{s\={o}l\={\i} lapid\={e}s}}

\def\apjl{ApJL }
\def\aj{AJ }
\def\apj{ApJ }

\def\apjs{ApJS }
\def\araa{ARAA }
\def\aap{A\&A }
\def\nat{Nature }
\def\mnras{MNRAS }
\def\na{New Astronomy}
\def\pasj{PASJ}
\def\actaa{ACTAA}
\def\icarus{Icarus}
\def\bain{Bulletin of the Astronomical Institutes of the Netherlands}
\begin{document}

%\def\thetitle{The characteristics of free-floating planets from an
%              Orion Trapezium-like star cluster}
%\def\thetitle{Survivability of planetary systems in young and dense star clusters}

%\usepackage{fontspec}
%\usepackage{pst-node}
%\setmonofont[Color={0019D4}]{Courier New}
\lstset{basicstyle=\ttfamily\small,breaklines=true, xleftmargin=1cm}
%\begin{document}

%\pagestyle{myheadings}

%%\author[1]{Arjen van Elteren}
%%\author[1 $\dag$]{Simon Portegies Zwart}
%%\author[2]{Inti Pelupessy}
%%\author[1]{Maxwell Cai}
%%\author[3]{Stephen L.W. McMillan}
%%
%%\affil[1]{
%%Leiden Observatory, Leiden University, PO Box 9513, 2300 RA, Leiden, The Netherlands
%%}
%%\affil[2]{
%%The Netherlands eScience Center, The Netherlands
%%}
%%\affil[3]{
%%Drexel University, Department of Physics and Astronomy, Philadelphia, PA 19104, USA
%%}

%% \affil[$\dag$]{E-mail: spz@strw.leidenuniv.nl}
%%%% Subject entries to be placed here %%%%
%\subject{xxxxx, xxxxx, xxxx}

%%%% Keyword entries to be placed here %%%%
%\keywords{xxxx, xxxx, xxxx}

\date{Accepted XXX. Received YYY; in original form ZZZ}

% for MNRAS:
% Enter the current year, for the copyright statements etc.
%\pubyear{2018}

\title{Survivability of planetary systems in young and dense star clusters}
\author {A. van Elteren
        \inst{1}
        \and
        S.\, Portegies Zwart\thanks{e-mail: spz@strw.leidenuniv.nl}\inst{1}
        \and
        I.\, Pelupessy \inst{2}
        \and
        Maxwell\, X.\, Cai \inst{1}
        \and
        S.L.W. McMillan \inst{3}}

        \institute{
         Leiden Observatory, Leiden University, NL-2300RA Leiden, The Netherlands
         \and
         The Netherlands eScience Center, The Netherlands
         \and
         Drexel University, Department of Physics, Philadelphia, PA 19104, USA
}

\abstract {}
{
  We perform a simulation using the Astrophysical Multipurpose Software
  Environment of the Orion Trapezium star cluster in which the
  evolution of the stars and the dynamics of planetary systems are
  taken into account.  
 }
  {The initial conditions from
  earlier simulations were selected in which the size and mass distributions of the
  observed circumstellar disks in this cluster are satisfactorily
  reproduced. Four, five, or size planets per star were introduced in orbit around the 500
  solar-like stars with a maximum orbital separation of 400\,au.
}
{
  Our study focuses on the production of free-floating planets.  A
  total of 357 become unbound from
  a total of 2522 planets in the initial conditions of the simulation. Of these, 281 leave the cluster within
  the crossing timescale of the star cluster; the others remain bound
  to the cluster as free-floating intra-cluster planets.  Five of
  these free-floating intra-cluster planets are captured at a later
  time by another star.
 }
 {
  The two main mechanisms by which planets are lost from their host
  star, ejection upon a strong encounter with another star or internal
  planetary scattering, drive the evaporation independent of planet
  mass of orbital separation at birth.  The effect of small
  perturbations due to slow changes in the cluster potential are
  important for the evolution of planetary systems.  In addition, the
  probability of a star to lose a planet is independent of the planet
  mass and independent of its initial orbital separation.  As a
  consequence, the mass distribution of free-floating planets is
  indistinguishable from the mass distribution of planets bound to
  their host star.
  }

%\footnotetext[0]{Submitted to: -}
%\begin{fmtext}
%\keywords{}
%\end{fmtext}
%\begin{multicols}{2}
%%%%%%%%%%%%%%% End of first page %%%%%%%%%%%%%%%%%%%%%

\maketitle

\section{Introduction}

In recent years several free-floating planets, i.e.,  planets not
orbiting a star, have been discovered by direct infrared imaging
\citep{2013ApJ...778L..42P} and bycatch in gravitational
microlensing surveys
\citep{2011Natur.473..349S,2012ARA&A..50..411G,2013ApJ...764..107G}.
Following star formation theory planets could in principle form in
isolation
\citep{2007AJ....133.1795G,2013ApJ...777L..20L,2015MNRAS.446.1098H},
but it seems more likely that they form according to the canonical
coagulation process in a disk orbiting a host star
\citep{Kant1755}. If planets are not formed in isolation, there are
three major mechanisms by which planets can be liberated.  A planet
may become unbound as a result of (\textit{i}) dynamical interaction with another star
    \citep{2002ApJ...565.1251H,2017A&A...608A.107V, 2017MNRAS.470.4337C, 2018MNRAS.474.5114C, 2015MNRAS.453.2759Z},
(\textit{ii\textup{)}}    scattering interactions among the planets in a
    multi-planet system \citep{2012MNRAS.421L.117V, 2017MNRAS.470.4337C, 2018MNRAS.474.5114C}, (\textit{iii})
  copious mass loss in a post-AGB phase
    \citep{2015MNRAS.451.2814V,2016RSOS....350571V}
or supernova
    explosion of the host star \citep{1961BAN....15..265B},  and (\textit{iv}) the ejection of fragments when the protoplanetary
    disk is perturbed \citep{2017A&A...608A.107V}.
The relative importance of each of these and other possible processes
are hard to assess, but the four listed here are probably most common.

A total of 20 free-floating planet candidates have been identified
\citep{2008AcA....58...69U,2010AJ....140.1868W,2015ARA&A..53..409W,2018arXiv181100441M}.
Two of these orbit each other in the binary-planet 2MASS~J11193254-1137466
\citep{2017ApJ...843L...4B}, but all others are single. Weak
micro-lensing searches indicate that the number of free-floating
planets with masses exceeding that of Jupiter is about one-quarter of the
number of main-sequence stars in the Milky Way Galaxy, whereas
Jupiter-mass planets appear to be twice as common as main-sequence
stars \citep{2011Natur.473..349S}.  Interestingly, Earth-mass
free floaters are estimated to be only comparable in number to
main-sequence stars \citep{2012Natur.481..167C}; there appears to be
a peak in the number of free-floating planets around the mass of
Jupiter.

If rogue planets are liberated upon a strong encounter with another
star in a cluster, this process is likely to take place during its
early evolution after circumstellar disks have coagulated into planets
and most of the primordial gas has been lost.  By this time, the
stellar density is still sufficiently high that strong encounters
between stars are common \citep{2015MNRAS.451..144P}. Young star
clusters may, therefore, make an important contribution to the
production of free-floating planets.  However, this is at odds with
the low number of free-floating planets seen in star clusters.  Only
one rogue planet was found in the TW Hydra association
\citep{2016ApJ...822L...1S} and a dozen candidates were found in the
sigma Orionis cluster \citep{2013MmSAI..84..926Z}, but no planets were
found in the Pleiades cluster despite active searches
\citep{2014A&A...568A..77Z}.  These estimates are in sharp contrast to
the number of asteroids and other \soli\footnote{\solus, means
  ``lonely rock'' in Latin.} expected from the star formation
processes \citep{2018MNRAS.479L..17P}.

The majority of free-floating planets appear as part of the field
population, but this may be a selection effect of the methods used to
find them \citep{2015ARA&A..53..409W}. To some degree, however, their
relatively high abundance in the field does not come as a surprise.
If every star that turns into a white dwarf liberates its planets (and
other debris), the number of isolated free floaters should exceed the
number of white dwarfs at least by the average number of planets per
star. Many of these stars are then already part of the field
population once they turn into white dwarfs, giving a natural
reduction of free-floating planets in clusters compared to the field
population. However, this would mean that dynamical interactions and
internal planetary instabilities have a minor contribution to the
formation of free-floating planets.

In order to investigate the consequences of stellar evolution and
dynamical interactions on the production of free-floating planets, we
perform a series of calculations in which we take the relevant
processes into account. The main question we address is to what degree
the dynamics of a star cluster contribute to the formation and variety
of free-floating planets, and what is the relative importance of the
various channels for producing these planets.

Planetary systems in our simulation are born stable in the sense that
allowing the systems to evolve in isolation would not result in
dynamical interactions among the planets.  This enables us to study
specifically the relative contribution of dynamical interactions on
the production of free-floating planets. The stars in our simulations
that receive a planetary system are selected such that they remain on
the main sequence for the entire duration of the simulation. Stellar
mass loss, therefore, does not specifically affect these planetary
systems. As a result, in the absence of dynamical interactions these
planetary systems are not expected to be affected by either internal
planetary dynamics nor by stellar mass loss.

We include, in our simulations, the gravitational interactions between
the stars, the interactions inside the planetary systems, and the mass
loss due to stellar evolution.  In principle, all the three main
processes mentioned above are included, although, as mentioned
earlier, the effect of stellar evolution is limited by the duration of
our simulations. We take all these effects into account as accurately
as our computer resources permit, which is particularly important for
the long-term dynamical processes among planets orbiting a single
star. The simulations are performed using the Astrophysical
Multipurpose Software Environment \citep[{\tt AMUSE};][]{2009NewA...14..369P,2013CoPhC.183..456P,2013A&A...557A..84P}. We
perform our calculations using a dedicated script, which we call {\tt
  Nemesis}, that enables us to integrate the equations of motion of
stars with planetary systems and includes the effects of mass
loss due to stellar evolution and collisions between stars and
planets. Our calculations ignore the primordial gas in the star
cluster, but our initial conditions are selected to mimic the initial
stellar and planet distribution functions shortly after the primordial
gas was expelled and the disks turned into planetary systems.  Several
example scripts of how {\tt AMUSE} operates and a more detailed
description of the framework is provided in \cite{AMUSE}.

In this work, we focus on the liberation processes and their
consequences in a dense star cluster with characteristics comparable
to the Orion Trapezium cluster. The majority of the observed field
stars and rogue planets may originate from bound clusters, loosely
bound associations, and only a minority from isolated stars.  Our
adopted initial conditions originate from a previous study
\citep{2016MNRAS.457..313P} in which the size distribution of
circumstellar disks in the Orion Trapezium cluster were reproduced.
We considered these conditions suitable for our follow-up study
assuming that some of the surviving disks would produce a planetary
system.  The cluster in the study of \cite{2016MNRAS.457..313P} was
born in virial equilibrium with a fractal density distribution with
dimension $F=1.6$.  The cluster initially contained 1500 stars with a
virial radius of 0.5\,pc.  At an age of 1\,Myr the size distribution
of the disks in this cluster is indistinguishable from the observed
size distribution of 95 ionized protoplanetary disks larger
than 100\,au in the Trapezium cluster \citep{2005A&A...441..195V}.

We adopt the earlier reconstructed initial parameters for the
Trapezium cluster and populate the stars that have a surviving disk with a
planetary system. The 500 stars with a disk size of at least 10\,au at
the end of their simulation received either four, five, or six planets with a
mean mass of $\sim 0.3\,M_{\rm Jupiter}$. The planets are assumed to
have circular orbits in a randomly oriented plane. The correlation
between orbital separation and planet mass was selected from the
oligarchic growth model for planetary systems by
\cite{2013ApJ...775...53H} and\cite{0004-637X-581-1-666}.

After the initialization, we continue the evolution of the star
cluster including its planetary systems for 10\,Myr to an age of
11\,Myr. At that time about half the cluster stars are unbound.

In the following section (\S\,\ref{Sect:Implementation}) we describe
the setup of our numerical experiment, followed by a description of
the initial conditions in \S\,\ref{Sect:ICs}.  We report on the
results in \S\,\ref{Sect:Results}, discuss the results in
\S\,\ref{Sect:Discussion} and eventually, in
\S\,\ref{Sect:Conclusions}, we summarize our findings.  In the
appendix (\S\,\ref{Sect:AppValidation}) we validate the adopted {\tt
  Nemesis} method for integrating planetary systems in stellar
clusters.

\section{Methods}\label{Sect:Implementation}

Integrating planetary systems in star clusters is complicated by the
wide range in timescales, ranging from days to millions of years, and
the wide range of masses, ranging from Earth-mass up to about 100\,\MSun.  The
first complication directly indicates that many planetary systems have
to be integrated over many orbits, which have to be realized without a
secular growth of the error in the energy. The wide range in masses
hinders such integrations by introducing round-off and integration
errors \citep{2015ComAC...2....2B}. The effect of stellar mass loss
complicates the numerical problem.  In this section, we describe the
methods developed to address these issues.

We use AMUSE for all the calculations presented in this work. This
framework is a component library with methods for coupling multi-scale
and multi-physics numerical solvers for stellar evolution,
gravitational dynamics, hydrodynamics, and radiative transfer. In this
paper we incorporate stellar evolution of all the stars in the
simulation via the {\tt SeBa} parametrized stellar evolution code
\citep{1996A&A...309..179P,1998A&A...332..173P,2012A&A...546A..70T,2012ascl.soft01003P}.
Gravitational interactions between planets are addressed using {\tt
  Huayno}, which is a class of a large variety of $N$-body codes based
on various kick-drift-kick algorithms via the Hamiltonian splitting
strategy of tunable order \citep{2012NewA...17..711P}. For this work,
we adopted the fourth and eighth order shared time-step solvers
\citep{1992PASJ...44..141M,2008NewA...13..498N}.  In our case, we
adopted the fourth order method for integrating the equations of
motion for the stars and the symplectic higher order method for
planetary systems.

The computing time for integrating Newtons' equations of motion of $N$
stars in a cluster scale $\propto N^2$. In a relatively small star
cluster such as the Trapezium cluster studied in this paper, the
integration time step for the top-level parent particles peaks at a
fraction of the mean cluster's crossing timescale, whereas the
planetary time step is typically on the order of a few percent of the
orbital period around the host star. A multi-time-step approach
consequently saves enormously in terms of computer time \citep[see
  also][]{1985mts..conf..377A}.

Adding planets to stars increases the number of particles in the
system. A more severe performance bottleneck is introduced by the
generally tight orbits in which these planets are introduced; i.e., years
for planets compared to millions of years for the free-floating stars
in the cluster. If all the new objects were introduced in a
regular $N$-body code the computation would come to a grinding halt.
To prevent this from happening and to reduce the effect of
integration errors and round off, we developed the {\tt Nemesis}
package within the {\tt AMUSE} framework.

The principles that make {\tt Nemesis} efficient is based on the wide
range of scales, which are used as an advantage by separately solving
systems that are well separated in terms of temporal or spatial
scales.  In addition, we introduce the simplification that a planet
orbiting one star has a negligible effect on the orbit of a planet
around another star in the cluster. This strict separation
subsequently allows us to choose different integrators for stars and
planetary systems. The latter flexibility allows us to tailor the
integration method to the topology of the system. As a consequence,
our calculations are naturally parallelized over the many
well-separated systems. This results in an enormous acceleration when
running on multiple cores because each of the $N$-body integrators can
run in parallel for the global intersystem communication timescale.
At the same time, energy is conserved per individual system and
separately for the global $N$-body system to machine precision.  This
combination of excellent performance and energy conservation makes
{\tt Nemesis} an ideal tool for integrating planetary systems in star
clusters.

\subsection{{\tt Nemesis} module}

In {\tt Nemesis}, planetary systems and stars are integrated together.
The underlying assumption is that the entire cluster can be separated
into groups.  We call these groups ``subsystems'' or ``children'' and
they can be composed of stars as well as planets that are relatively
close together with respect to the size of the cluster.  The dynamics
in these subsystems is not resolved in the global integrator, which we
call the ``parent'', but is integrated separately.  In many cases, a
planetary system is a subsystem, but children may also be composed of
several planetary systems that happen to be spatially in close
proximity.  In this approach, we integrate subsystems separately from
the rest of the cluster, but the components of the subsystems and the
other cluster objects feel each other's forces.

\subsubsection{Calculating forces}

In this section, we explain how the forces in the {\tt Nemesis} module
are calculated.  To ease the discussion, we define the term particle.
Particles represent the center of masses of a subsystem or of
individual objects, such as single stars or free-floating planets.
Particles represent the parents in the $N$-body system and are
integrated together in one $N$-body code. In practice, the particles
are integrated with a fourth- or sixth-order Hermite predictor-corrector
method \citep{1992PASJ...44..141M,2008NewA...13..498N}.

The internal dynamics of each child (the subsystem) is integrated with
a separate N-body code. The latter can be a different code, for
example, a simple Kepler solver or some high-order symplectic N-body
solver. We call this the local subsystem for a particular particle, or
the parent's child.  The entire simulation is then composed of as many
N-body codes as there are subsystems and one additional code for all
the particles that are not part of a subsystem including the center of
masses of all the subsystems. The parent system is then composed of
subsystems, single stars, or planets.

\begin{figure*}[htb]
\centering
\includegraphics[width=0.95\textwidth]{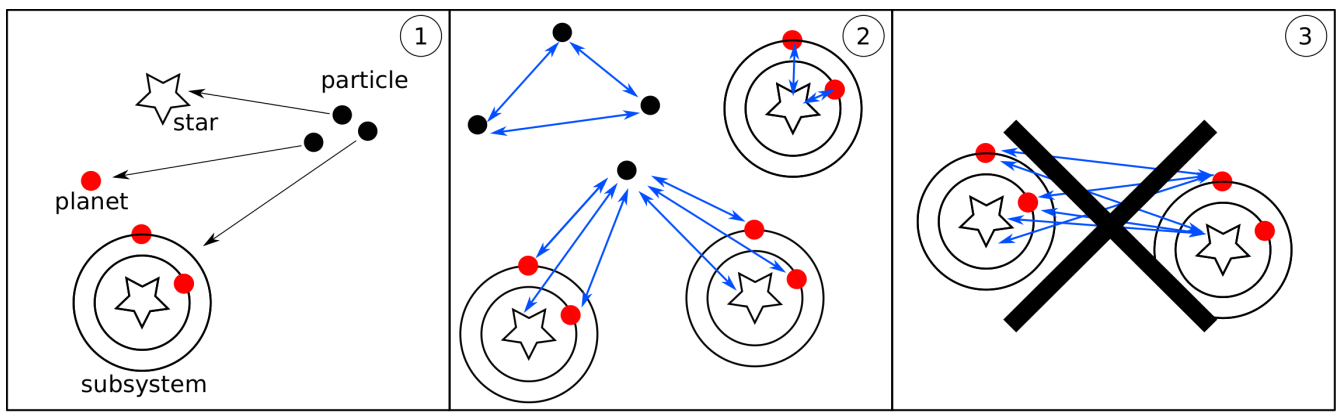}
\caption{{Diagram of \tt Nemesis} method. (A) A particle is an individual
  object or a subsystem consisting of multiple individual objects. In
  this study, the individual objects are either stars or planets. (B)
  The gravitational force on a particle is the sum of the force from
  other particles [1] and the forces from the individual objects in
  the subsystems. (C) The gravitational force on individual objects is
  the sum of the force from the particles [1] and the forces from the
  other individual objects in the containing subsystem [2], but not
  from those individual objects in other subsystems [3]. The forces
  from A.1 and B.2 are each in a self-contained system and can be
  calculated in an $N$-body code; the forces in A.2 and B.2 are
  connected to the self-contained systems and are evolved with a
  leapfrog algorithm.}
\label{fig:doodle1}
\end{figure*}

The gravitational force exerted on each particle is composed of three
parts:
the forces from all the other objects in the local
    subsystem,
   the forces of all the single particles in the global
    system,   and the force of the stars and planets in the
    other subsystems.
In {\tt Nemesis} we ignore the forces of the individual objects
(planets and stars) in the other subsystems. Instead, we take the force
from the center of mass of the subsystem into account.  As a
consequence the stars and planets in a subsystem feel the total
force from other subsystems as exerted from the center of mass of that
subsystem, but not the individual forces from all the individual
components from within that subsystem. Particles in other subsystems, therefore, do not feel the forces of individual planets orbiting a star
in the other subsystem.  Local particles feel the forces of the
other planets and stars in the same system. This procedure, outlined
in figure \ref{fig:doodle1}, results in a slight error in magnitude
and direction of the force on any particle due to the assumption that all
objects in another subsystem exert a force from the center-of-mass of
that subsystem. As long as a subsystem is composed of a star with some
planets, this error remains small, but the error grows when a
subsystem is composed of multiple stars. We reduce this error by
assuring that subsystems remain small compared to the interparticle
distance and that they are not composed of many stars.

\subsubsection{Integrating the system}

The force calculation in Nemesis is implemented in multiple bridge
operations \citep{2007PASJ...59.1095F,AMUSE}.  These bridges integrate
the equations of motion of the individual components (particles and
the subsystems) via a second-order Verlet kick-drift-kick method
\citep[see][]{1995ApJ...443L..93H,2014A&A...570A..20J}.

In the initial kick phase, we accumulate the forces between the single
particles and the particles in each of the subsystems. These forces
are used to update the velocities of the particles and the objects in
each of the subsystems over half a bridge time step, $dt_{\rm
  bridge}/2$.

In the drift phase, the particles and subsystems are integrated using
the forces between the particles in each individual subsystem. Since
this is an uncoupled problem, each individual subsystem is integrated
in parallel. In this phase, we ignore the forces between the single
particles and those that are in subsystems.
In the final kick phase, we again calculate the forces between the 
single particles and the particles in the subsystems based on the new
positions after the drift phase, and again update the velocities.

This procedure allows us to integrate particles and subsystems
independently. This strict separation of integrating subsystems
enables us to adopt a different N-body code for each subsystem,
although this is not a requirement. In addition, it makes the
concurrent integration of each subsystem possible, which enormously
speeds up the procedure for a sufficiently large number of subsystems.

\subsubsection{Subsystem dynamics}

Subsystems may change their composition at runtime. This can happen
when a star or planet is ejected, planets or stars collide, when two
or more subsystems merge, or when a single object enters the
subsystem.  To simplify this process, we recognize two changes to a
subsystem:
\begin{description}
\item[\textbf{Merger}] Two subsystems are merged to one as soon as
  their center of masses approaches each other to within the sum of
  their radii.  In this case the radius of each subsystem is the maximum of
  two radii: it is (1) 5\% larger than the distance from the center-of-mass to the outer-most object and (2) the size that corresponds to a
  likely encounter. The latter is a function of the bridge time step
  ($t_{\rm nemesis}$), the number of objects in a subsystem, the mass
  of the subsystem, and a dimensionless factor $\eta$: $t_{\rm enc} =
  1.0 / \eta t_{\rm nemesis}$. We adopt a value of $\eta \simeq
  0.2$. Upon the merger of two subsystems, one of the N-body
  integrators assimilate the other subsystem and the other
  integrator is terminated.  Since both integrators may be different,
  we assume that the integrator with the largest number of particle
  survives.

\item[\textbf{Dissolution}] A subsystem can dissolve into individual
  objects or multiple body parts can split off to form their own
  separate subsystem.  The procedure to decide on the dissolution of a
  subsystem follows the inverse criteria as for the merger of two or
  more subsystems.  This procedure may lead to the starting of one or
  more new integrators to take care of the various newly introduced
  subsystems.  Single objects (stars or planets) are incorporated in
  the global integrator when they escape from a subsystem.

\end{description}
From an astronomical point of view, this procedure looks somewhat
arcane, but numerically it has many advantages because it allows us to
optimize for efficiency, performance, and accuracy.

\subsubsection{Planet and stellar collisions}

Apart from the dissolution and merging of dynamical subsystems, we
also allow stars and planets to experience physical collisions.
Collision can only occur within a subsystem.  If two stars in the
parent system were to collide, they would first for a separate
subsystem within which the collision is handled.  Two stars or
planets are considered to collide as soon as their mutual distance is
smaller than the sum of their radii.  A collision always results in a
single object, while conserving the mass, volume, and angular momentum
in the collision. In principle, it would be relatively easy to perform
a hydrodynamics simulation upon each collision, but that is beyond
the scope of our current study. A more extensive discussion on
such more rewarding events is provided in \cite{AMUSE}.

Isolated stars have a size according to the stellar evolution code,
which runs concurrently with the dynamics.  The sizes of planets are
calculated by assuming a mean planet density of $3$\,g/cm$^3$.  For
improved efficiency, we adopt a special treatment for collisions
between planets and the central star of a planetary system.  Planets
are assumed to collide with their orbiting star as soon as they
approach it to within 1\,au.  This relatively large distance was
adopted in order to reduce the computational cost of integrating tight
planetary orbits and to minimize the errors associated with their
numerical integration. We can easily relax this assumption, but it
would result in a considerable increase in computer time.

The new mass of a merged object is the sum of the two individual
masses and the new position and velocity are determined by conserving
linear momentum and angular momentum. The radius of the collision product
of two planets is calculated by conserving the density. A stellar
collision acquires its new radius on the stellar evolution track as
described in \cite{1996A&A...309..179P}.

\subsection{Selecting the $N$-body codes in Nemesis}

Each subsystem is integrated with a separate $N$-body code. In
principle, each of these codes could be different. In practice,
however, we use two different techniques to integrate the equations of
motion of the stars and planets.  The choice of code is based on the
requirements for the physics.

For two-body encounters, we adopt a semi-analytic Kepler solver as
implemented by \cite{2013MNRAS.429..895P}.  For a typical planetary
system in which one particle is much more massive (at least more than
100 times) than the other particles, we use {\tt
  Rebound} \citep{2012A&A...537A.128R} with an implementation of a
symplectic Wisdom-Holman integrator
\citep[WHFAST][]{2015MNRAS.452..376R}.  For all other subsystems, we
adopt the eighth-order method available in the symplectic integrator {\tt
  Huayno} \citep{2012NewA...17..711P}.  The center-of-masses of the
subsystems, the single stars, and the free-floating planets are
integrated via the Hermite fourth-order predictor-corrector integrator
\citep{1992PASJ...44..141M,2008NewA...13..498N}.

All calculations are executed on a central processing unit (CPU) because the number of particle in
each N-body code is relatively small and a graphics processing unit
(GPU) would not provide many benefits in terms of speed
\citep{2008NewA...13..103B}.

\subsection{Validation and verification}

The performance and accuracy of the {\tt Nemesis} integrator module is
controlled with two parameters: one controls the distance for which
individual objects (planets and stars) and subsystems merge or
dissolved, and the another controls the time step of the bridge
operator. This so-called bridge time step controls the numerical
timescale for the interactions between the subsystems and the
particles.  Both parameters are tuned independently but we choose to
express the bridge time step in terms of the encounter distance and
the mass of the objects. This adopted scaling leaves only the {\tt
  Nemesis} time step, $dt_{\tt Nemesis}$, as a free parameter for
integrating the entire $N$-body system.  This timescale depends on
the topology of the $N$-body system, and we tune its value by
performing scaling and validation tests. A detailed analysis of the
dependency of the model on the time step in presented in Appendix
\ref{Sect:AppValidation}. For our choice of initial conditions and
integrators we found that an interaction time step of 100\,yr gives
the most satisfactory results in terms of reproducibility,
consistency, energy conservation, and speed.

\section{Initial conditions}\label{Sect:ICs}

After developing and validating the numerical framework we can start
generating the initial realization for our star cluster with planetary
systems.  We start the calculations with a cluster of stars, some of
which have a planetary system.  The initial realization is motivated
by \cite{2016MNRAS.457..313P}, who studied the dynamical evolution of
the star cluster with 500 to 2500 stars taken from a broken power-law
mass function between 0.1\,\MSun\, and 100\,\MSun\,
\citep{2001MNRAS.322..231K}. These calculations were performed with a
fourth order Hermite $N$-body method including a heuristic description
for the size and mass evolution of circumstellar disks.  At the start
of these calculations each star received a disk with a mass of 1\% of
the stellar mass and a size of 400\,au. During the $N$-body
integration the sizes and masses of these disks were affected by close
stellar encounters \citep{2016MNRAS.457.4218J}.  During these
simulations the disk size distributions were compared with the
protoplanetary disks observed using Hubble Space Telescope WFPC2 of
the Trapezium cluster \citep{2005A&A...441..195V}. In this way
\cite{2016MNRAS.457..313P} was able to constrain the initial cluster
parameters.  Clusters for which the stars were initially distributed
according to a \cite{1911MNRAS..71..460P} distribution did not
satisfactorily reproduce the observed disk-size distribution,
irrespective of the other parameters, but when the stars were
initially distributed according to a fractal with a dimension $F=1.6$
and in virial equilibrium ($Q=0.5$) the simulations satisfactorily
reproduced the observed disk size distribution in the Trapezium
cluster (KS probability of $\sim 0.8$) in the age range from 0.3\,Myr
to 1.0\,Myr. For our simulations, we adopted the final stellar masses,
positions, and velocities for one of these simulations that matched
the observed distribution of disk sizes and disk masses best. As a
consequence, our initial conditions had already evolved dynamically
for 1\,Myr before we started our calculation.

In Table\,\ref{Tab:SPZ2016} we present the initial parameters as
adopted by \cite{2016MNRAS.457..313P} in the left column (indicated by
$t = 0$\,Myr). The third column gives the global cluster parameters at
an age of 1\,Myr, which are the final conditions for the study
performed by \cite{2016MNRAS.457..313P}.  We adopted these parameters and
in fact, the precise realization of these calculations as initial
conditions for our follow-up calculations.  The last (fourth) column
presents the global cluster parameters at the end of our
simulations, at an age of 11\,Myr, which is 10\,Myr after the
introduction of the planetary systems.

\begin{table}
  \caption{Initial cluster model adopted by
    \cite{2016MNRAS.457..313P}; the final conditions for the disk-size
    analysis in \cite{2016MNRAS.457..313P}, which we adopted as
 the    initial realization for the simulations presented here; and the
    final conditions. The parameter $N_{\rm total}$ indicates the total number of stars in the 
    simulation;  $N_{\rm bnd}$, the bound mass of the cluster in solar masses; $R_{\rm vir}$
    the virial radius of the cluster;  $Q$, the virial equilibrium; $F$, the 
    fractal dimension of the cluster; $N_{\rm bnd}$, the number of bound
    stars; $N_{\rm bnd}$, the number of bound stars with planets; 
    $N_{\rm bnd}$, the number of unbound stars with planets; $N_{\rm w/planets}$,
    the number of stars with planets; $N_{r \geq 100au}$ the number of stars with
    disks or planets equal or larger than 100 au; $N_{r \geq 100au}$ the number of stars with
    disks or planets equal or larger than 10 au; $n_{\rm bnd}$, the number of planets
    bound to a star; $n_{\rm ff}$, the number of free-floating planets;    $n_{\rm unbnd}$, the number of planets unbound from the cluster;  
    $m_{\rm bnd}$, the total planetary mass in Jupiter masses, bound to a star; $m_{\rm ff}$, the total
    planetary mass in free-floating planets; and $m_{\rm unbnd}$, the total planetary mass unbound from the cluster. 
 }
  \label{Tab:SPZ2016}
  \begin{center}
  \begin{tabular}{llll}
    \hline
Parameter & $t=0$\,Myr & $t=1$\,Myr & $t=11$\,Myr\\
\hline
\multicolumn{4}{l}{Cluster characteristics} \\
  $N_{\rm total}$    & 1500 & 1500 & 1482 \\
  $M_{\rm bnd}$/\MSun & 627 & 618 & 545 \\
  $R_{\rm vir}$/pc  & 0.5  & 0.36 & 0.32 \\
  Q                & 1.0  & 0.6 & 1.0 \\
  F                & 1.6 & 1.26 & 0.6 \\
\multicolumn{4}{l}{Stellar characteristics} \\
  $N_{\rm bnd}$      & 1500 & 977 & 508 \\
  $N_{\rm bnd,w/p}$   &   0 &  512 & 166 \\ 
  $N_{\rm unbnd,w/p}$ &   0 &   0 & 323 \\
  $N_{\rm w/planets}$ &    0 & 500 & 517 \\
  $N_{r \geq 100au}$  & 1500 & 78 \\
  $N_{r \geq 10au}$   & 1500 & 578 \\
\multicolumn{4}{l}{Planets characteristics} \\
  $n_{\rm bnd}$      &    -- & 2522 & 2165 \\
  $n_{\rm ff}$       &    -- &   0 & 357 \\
  $n_{\rm unbnd}$     &    -- &   0 & 282 \\
  $m_{\rm bnd}$/\MJup &    -- & 3527 & 2915\\
  $m_{\rm ff}$/\MJup  &    -- &  0 & 502\\
  $m_{\rm unbnd}$/\MJup &   -- & 0 & 395\\
\hline
\hline
  \end{tabular}
  \end{center}
\end{table}

During the first 1\,Myr of evolution, starting from a fractal spatial
distribution (see the leftmost panel in Fig.\,\ref{fig:initial_positions})
most of the structure in the initial cluster is lost. The cluster seems
to have expanded considerably, as is evidenced by the zoom-out in
Fig.\,\ref{fig:initial_positions}, but when considering the virial
radius has in fact decreased from the initial 0.5\,pc to $R_{\rm vir}
\simeq 0.36$\,pc at an age of 1\,Myr.  At this moment, we randomly
select 500 stars for which the circumstellar disk has survived with a
radius of at least 100\,au. We subsequently assign a planetary system
to 500 of the stars with a surviving disk. The total mass of the
planets is identical to the disk mass. The masses and orbital
separation of planets are generated using the oligarchic growth model
\citep{1998Icar..131..171K} between a distance of 10\,au\, to 400\,au
from the host star.

There is no particular reason why we adopted a minimum separation of
10\,au, but adopting a smaller minimum separation would have resulted
in many more planets with a low mass in very tight orbits. This would
have resulted in an enormous increase in computing time.  All
planets have initially circular orbits with inclination randomly
selected from a Gaussian distribution with a dispersion of
1$\degree$\, around a plane.  This plane is defined as the
orbital plane of the planet closest to the star.  After the planetary
systems are initialized they are rotated to a random isotropic
orientation. Each star acquires between 4 and 6 planets with a mass of
0.01 to 130 Jupiter masses (see Fig.\,\ref{fig:initial_hist} and
Fig.\,\ref{fig:initial_meanmass}).  The total number of planets in the
simulation was 2522.

\section{Results}\label{Sect:Results}

When starting the simulation the stars are already 1\,Myr old and the
stellar density and velocity distribution are the result of the
previous calculations reported in \cite{2016MNRAS.457..313P}.  We
continue to evolve this cluster including its planets for 10\,Myr to
an age of 11\,Myr.

We performed one simulation in which all interactions between stars and
planet are taken into account using {\tt Nemesis}. Snapshots are
produced every 1000\,yr, but most of the analysis aims at the final
snapshot at an age of 11\,Myr.  A second simulation was performed in
which the planetary systems are evolved in isolation without any
interactions from other stars. This second run is used for validation
purposes only. Even though not explicitly discussed, no free-floating
planets were formed in this second run because the initial planetary
configurations are intrinsically stable.

\subsection{Global evolution of the star cluster}

\begin{figure*} %%[!htb]
\centering
\hspace*{-0.5cm}
\includegraphics[width=1.1\textwidth]{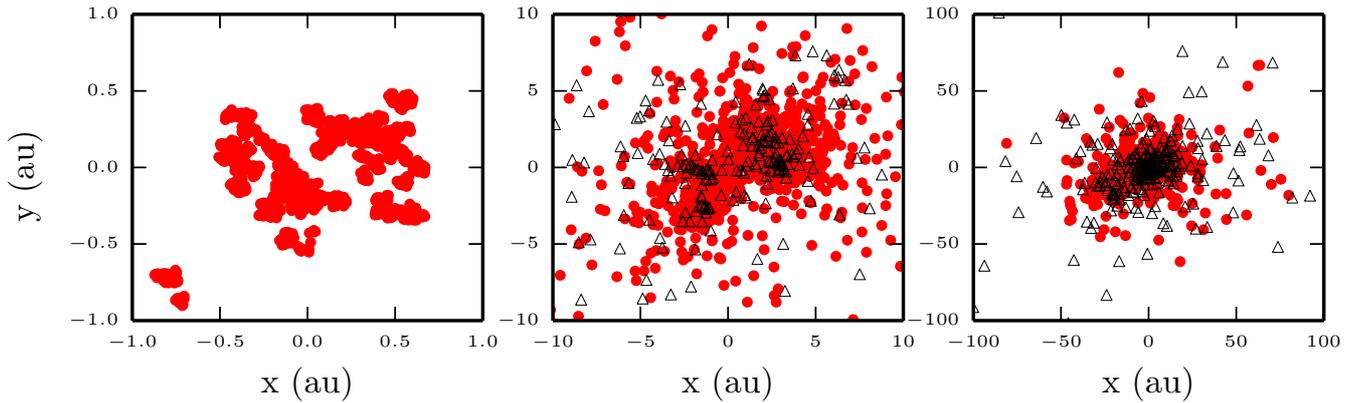}
\caption{Projected view of the simulated star cluster at $t=0$\,Myr
  \citep[initial conditions adopted by][]{2016MNRAS.457..313P} (left
  panel), at $t=1$\,Myr (middle panel and the adopted initial
  conditions), and at $t=11$\,Myr (right panel, our final
  conditions). Stars are red bullets, single free floating planets
  black triangles. 
}
\label{fig:initial_positions}
\end{figure*}

In Fig.\,\ref{fig:initial_positions} we present a projected view of
the stars and planets of our simulated cluster at birth (left), at an
age of 1\,Myr (middle) and at the end of the simulation, at an age of
11\,Myr.
During the first 1\,Myr in which the stars still have circumstellar disks the
cluster loses most of its initial fractal structure.  During this
early phase, the cluster is most dynamically active and the majority
of stars experience one or more close encounters with other
stars. These encounters cause the truncation of circumstellar disks.
By the time we introduce the planetary systems, at an age of 1\,Myr,
most dynamical interactions have subsided and the cluster has expanded
by about an order of magnitude, although the cluster core remains
rather compact (see also Table\,\ref{Tab:SPZ2016}). The reduction in
density has profound consequences for the survivability of our
planetary systems.  During the subsequent 10\,Myr of evolution the
outer parts of the cluster expand by another order of magnitude, but
the cluster core remains rather small and bound.

In the overview presented in Table\,\ref{Tab:SPZ2016} we demonstrate
that the cluster hardly loses any mass during its evolution.  Mass
loss due to stellar winds is rather moderate, reducing the total
cluster mass from 618\,\MSun\, to 545\,\MSun\, in 10\,Myr.  The
majority of this mass loss is caused by the two most massive stars of
73\,\MSun\, and 64\,\MSun. These stars experience copious mass loss in
the Wolf-Rayet phase followed by a supernova explosion.  Such
evolution may enrich most of the disk in the cluster by r-processed
elements \citep{2018A&A...616A..85P}.  The expansion of the cluster by
about an order of magnitude and the global mass loss in bound stars
cannot be attributed to the stellar mass loss alone. In total, the
cluster loses about two-thirds of its stars, one-third in the first
Myr, and another third in the following 10\,Myr.  The structure of the
cluster also changes from an initial fractal dimension of $F=1.6$ to
$F=1.26$ at 1\,Myr and to $F\simeq 0.6$ at the end of the simulation.
The eventual cluster, at an age of 11\,Myr, can be well described with
a Plummer distribution \citep{1911MNRAS..71..460P} with a
characteristic radius of 0.32\,pc. Although, in
fig.\,\ref{fig:initial_positions} the cluster appears to expand by two
orders of magnitude, the cluster central portion remains rather
confined within a parsec.

\subsection{Characteristics of the surviving planetary systems}

During our calculations, planetary orbits are affected in a number of
ways. We start by describing the characteristics of the surviving
planetary systems. Later, in \S\,\ref{Sect:Collisions} and \S\,\ref
{Sect:FFPs} we discuss the planets that are lost due to collisions or
ejection from their host star.

\begin{figure}[!htb]
  \centering
    \scalebox{1.0}{
%Added by TeX Support
      \input{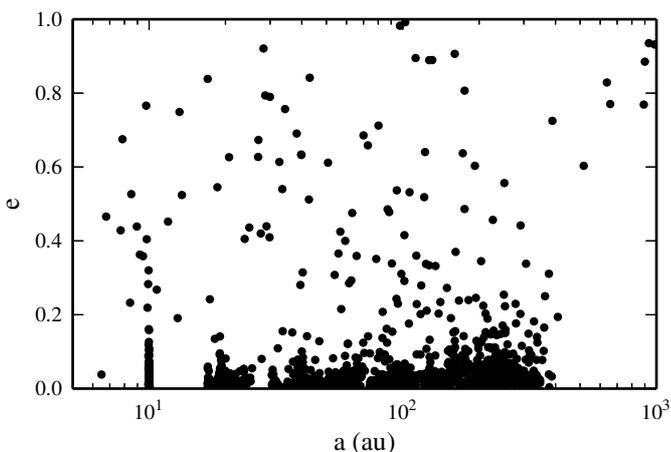}
      }
\caption{Eccentricity as a function of the semimajor axis the planets
  that survive up to an age of $t=11$\,Myr.}
\label{fig:ecc_vs_semimajoraxis}
\end{figure}

In Fig.\,\ref{fig:ecc_vs_semimajoraxis} we present the distribution 
in eccentricity and semimajor axis of the planets that remain bound 
up to an age of 11\,Myr.  About 10\% (213 in total) of the planets 
have experienced considerable orbital variations ($\Delta e>0.1$ or 
$\Delta a>10\%$) due to a combination of encounters with other stars 
and internal planetary scattering. We note that in the absence of 
stellar encounters the planetary systems are not affected by 
internal scattering. Any changes in the planetary systems in our 
simulation is therefore the result of interactions with external 
perturbators (stellar encounters and cluster topology). These 
interactions put the planets in orbits where internal scattering 
causes further changes in the orbital parameters.

Some planets acquire eccentricities close to unity, indicating that
they may be subject to tidal interactions or even collisions with the
host star. Although we ignore tidal effects in our calculations,
collisions are taken into account.  A total of 75
($\sim 3.0$\,\% of the total) planets collided with their parent star
and 14 ($\sim 0.6\%$) planets experienced a collision with another
planet. We discuss planetary collisions more extensively in
\S\,\ref{Sect:Collisions}.

\begin{figure}[!htb]
\centering
\hspace*{-4mm}  
\includegraphics[width=1.15\columnwidth]{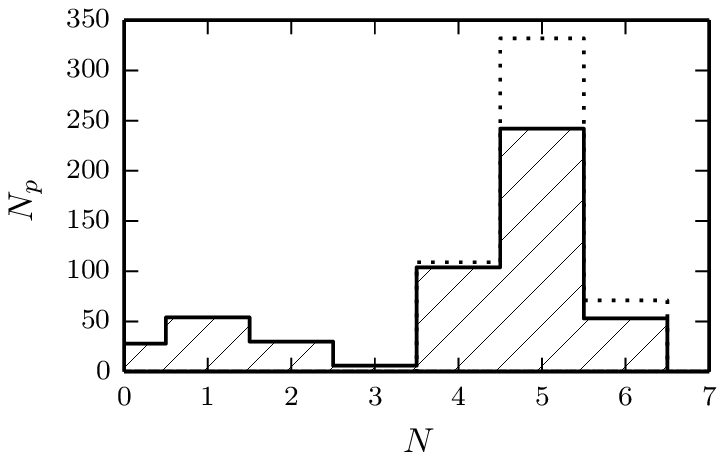}
\caption{Histogram for the number of systems with a certain number of
  planets.  The dotted curve gives the initial distribution with
  either 4, 5, or 6 planets per star.  The final conditions for the
  simulation without stellar dynamics are identical to this initial
  distribution.  The distribution of the simulation in which we
  included the stellar encounters at an age of 11\,Myr is presented as
  the solid curve with slanting lines.
%% Many planetary systems are reduced (in the sense of
%% having lost one or more planets) as a result of encounters, but
%% there is a dearth of systems with 3 planets.
}
\label{fig:initial_hist}
\end{figure}

In Fig.\,\ref{fig:initial_hist} we compare the distributions of the
number of planets per star in our simulation and compare the
distribution with the simulation in which we ignored any stellar
encounters. In the latter simulations, the planetary systems are not
affected by dynamics and their conditions remain very close to the
initial conditions.  This indicates that the initial configuration of
our planetary systems is stable against internal dynamical
evolution.

All stars with planetary systems have either four, five, or six planets
initially.  In Fig.\,\ref{fig:initial_hist} we subsequently observe
that in particular systems with five planets tend to be reduced, whereas
only a few stars with four planets or six planets seem to lose any. In
addition, by the end of the calculations, the number of systems with three
planets seem to be rather small compared to the number of systems with
one or two planets.  To further quantify the results we also present
Table\,\ref{Tab:planetsperstar}, in which we present the number of
planets for a star initially (columns) versus the final number of
planets (rows).

\begin{table*}
  \caption{Comparison of the distribution of planets at the beginning
    and at the end of the simulation.  In each cell, the count of the
    number systems with a certain number of planets is given. This
    count is given per number of planets in the original system. The
    original distribution 1\,$Myr$ is given in the top summation row; the
    final distribution at 11\,$Myr$ is given in the last column. For
    the final distribution of the 6 systems with 3 planets, 2 of these
    systems originally had 4 planets, 4 originally had 5 planets, and
    none originally had 6 planets. A total of 5 new planetary systems
    have been created during the evolution of the cluster, in these
    systems originally the star had no planets. Of these 5 new
    planetary system, 3 systems have 1 planet and 2 have gained 2
    planets.  }
  \label{Tab:planetsperstar}
  \begin{center}
  \begin{tabular}{l*{8}{>{\raggedleft\arraybackslash}p{1cm}}}\toprule
  & \multicolumn{7}{c}{$N_P$} & \\ \cmidrule(r){2-8}
    & 0  & 1  & 2  & 3  & 4  &  5 & 6 & $\sum$ \\
    \cmidrule(r){1-8}
    $\sum$&  5 &  0 &  0 &  0 &  109 & 332&  71& \\
    \cmidrule(r){2-8}
0 &  0 &  0 &  0 &  0 &  3 &  25 &  0 & 28 \\
1 &  3 &  0 &  0 &  0 &  6 &  35 &  10  &   54 \\
2 &  2 &  0 &  0 &  0 &  5 &  17 &  6  &  30\\
3 &  0 &  0 &  0 &  0 &  2 &  4 &  0  &   6\\
4 &  0 &  0 &  0 &  0 &  93 &  11 &  0  &   104\\
5 &  0 &  0 &  0 &  0 &  0 &  240 &  2  &  242\\
6 &  0 &  0 &  0 &  0 &  0 &  0 &  53 &    53 \\
\bottomrule
\end{tabular}
\end{center}
\end{table*}

From Table\,\ref{Tab:planetsperstar} we see that the systems with three
planets by the end of the simulation tend to originate from systems
with initially four or five planets. But we find that most systems that initially have five
planets reduce directly to one or no planets at all.  Curiously enough
though, systems that initially have six planets do not lose as many planets,
but when they do, they tend to reduce to a single planet, whereas for
systems that initially have four planets tend to be rather agnostic about how
many planets they lose. Statistically, these changes are significant
but much can be attributed to the initial conditions.  According to
our initial conditions, large disks with a relatively high mass are
prone to receiving more planets than small low-mass disks. The large
disks tend to be hosted by relatively low-mass stars, and those stars
tend to avoid the cluster center, whereas relatively high-mass stars
tend to be more abundant in the cluster core. These differences
propagate in the distribution of planets and therefore cause an
imprint on their future scattering history.

\begin{figure}[!htb]
\centering
%Added by TeX Support
\input{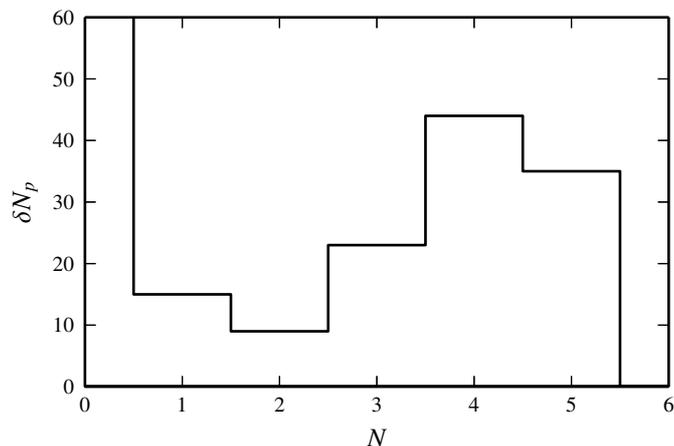}
\caption{Histogram of the number of systems with a certain number of
  lost planets. Of the original 500 planetary systems the majority
  (386) do not lose any planets. Only 25 systems lose 1 or 2 planets
  and 102 systems lose 3 or more planets. }
\label{fig:delta_hist}
\end{figure}

In Fig.\,\ref{fig:delta_hist} we plot the number of planets in a
planetary system at the end of our simulations. The majority of stars
keep all their planets throughout the calculations, but if a star
loses planets, it tends to lose a larger number like three to five rather than
just one or two.  The lost planets become free-floating or rogue planets,
which we discuss in \S\,\ref{Sect:FFPs}.

\begin{figure}[!htb]
  \centering
    \scalebox{1.0}{
%Added by TeX Support
      \input{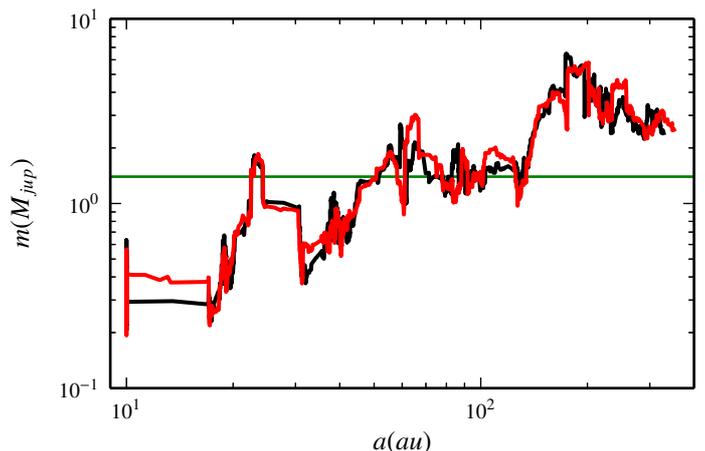}
      }
\caption{Mean planet mass as a function of semi major axis (in a
  moving bin of 50 planets). The initial (at 1\,Myr, in black) 
  and the final (at 11\,Myr in red) mean mass only differ slightly.
  The mean mass, 1.4 M$_{\rm Jupiter}$, is depicted with a green horizontal line.}
\label{fig:initial_meanmass}
\end{figure}

The redistribution of planets among the stars may also be affected by
the masses of the planets. To quantify this we present in
Fig.\,\ref{fig:initial_meanmass} the mean planet-mass as a function of
their semimajor axis.  The oligarchic-growth model, used to generate
the initial planetary systems, leads to more massive planets at larger
orbital separation (visible in Fig.\,\ref{fig:initial_meanmass}).  To
see if there is a mass preference for ejecting planets we also show,
in Fig.\,\ref{fig:initial_meanmass}, the final distribution (at an age
of 11\,Myr). Although the differences between both distributions
appear small, the differences at small separation are statistically
significant.

\begin{figure}[!htb]
  \centering
  \scalebox{1.0}{
%Added by TeX Support
    \input{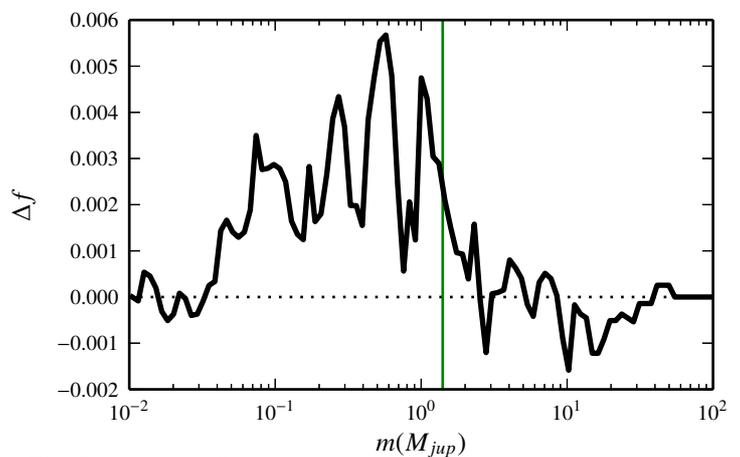}
    }
\caption{Relative difference between the cumulative distributes of the
  masses of the bound planets initially and at 11\,Myr. Positive
  values indicate an overabundance at the end of the
  simulation. Initially, the mean planet mass is $1.398 \pm
  1.05$\,\(M_{\rm Jup}\) at 11\,Myr the mean mass is only fractionally
  different at $1.404 \pm 4.191$ \(M_{\rm Jup}\); the latter value is
  indicated by the vertical green line.}
\label{fig:initial_planet_mass_hist}
\end{figure}

To further quantify these findings we present in
Fig.\,\ref{fig:initial_planet_mass_hist} the difference in the
cumulative distribution of planet mass for the cluster at an age of
1\,Myr with respect to 11\,Myr. The difference between the two
cumulative distributions are small and the fluctuations rather large,
but in the final systems, low-mass planets are more abundant than
high-mass planets. The turnover occurs near the mean-planet mass in
our simulation which is around $1.4$\,M$_{\rm Jupiter}$ (indicated
with the vertical line in Fig.\,\ref{fig:initial_planet_mass_hist}).
Based on the lack of a correlation between planet mass and orbital
separation we argue that the majority of ejections is driven by
external perturbations (mostly with other stars) rather than by
internal scattering among the planets.

\subsection{Migrating and abducted planets}\label{Sect:Captures}

Two rather extreme processes that affect the orbits of planets are
their abduction from another star or when a planet is scattered during a
close encounter with other planets. In both cases the resulting
planet is expected to be parked in a wide orbit with high
eccentricity. However, planets that are scattered close to the host star
into a parking orbit are expected to have higher eccentricity, on
average, than planets abducted from another star
\citep[see][]{2016MNRAS.457.4218J}.

\begin{figure}[!htb]
  \centering
  \scalebox{1.0}{
%Added by TeX Support
    \input{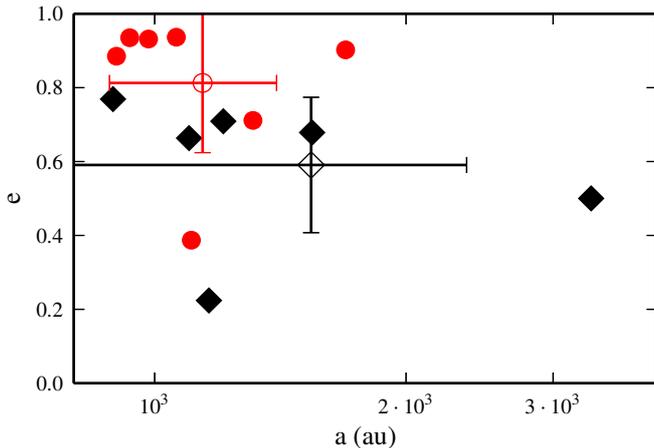}
    }
\caption{Eccentricity as a function of the semimajor axis for captured
  planets (black diamonds) and migrated planets with semimajor axis
  larger than 800\,au (red dots) at an age of $t=11$\,Myr. The mean
  and standard deviation for both sets are also plotted. The mean
  orbital elements for the captured planets is $a_c=1539\pm824$\,au
  and $e_c=0.6\pm0.2$, and $a_c=1141\pm258$\,au and $e_c=0.8\pm0.2$
  for the migrated planets. }
\label{fig:captured-vs-migrated}
\end{figure}

In our simulations, only a few planets were abducted, and a comparable
number of planets were kicked out to the outskirts of their own
planetary system by internal scattering. In
Fig.\,\ref{fig:captured-vs-migrated} we compare the orbital separation
and eccentricity of these systems.  Although the distributions are
rather broad in semimajor axis and in eccentricity, captured planets
have on average lower eccentricity and somewhat larger orbital
separation compared to ejected planets.

In table\,\ref{Tab:migratedplanets} we list the migrated planets, and
the abducted planets are presented in
Table\,\ref{Tab:capturedplanets}. Apart from slight differences in the
orbital parameters, the mass of the host star for captured planets
tends to be considerably higher than for the migrated planets. This
trend is not unexpected because of the stronger gravitational
influence of more massive stars whereas low-mass stars are more prone
to lose planets.

\begin{table}
  \caption{Parameters for planetary systems in which one planet was
    ejected to a larger distance ($>800$\,au) from its host star.  The
    second column identified which planet was ejected, followed by the
    mass of the host, planet mass, and its eventual orbital
    parameters.  }
\label{Tab:migratedplanets}
\begin{center}
\begin{tabular}{ll*{4}{>{\raggedleft\arraybackslash}p{1cm}}r}\toprule
 % & \multicolumn{7}{c}{$N_P$} & \\ \cmidrule(r){2-8}
  System & Planet & M $(M_{sun})$ & m $(M_{jup})$ & a (au) & e \\
  \hline
0 & a & 0.33 & 1.27 & 983.2 & 0.93 \\
1 & a & 0.16 & 2.23 & 900.1 & 0.89 \\
2 & a & 0.37 & 0.85 & 1107 & 0.39 \\
3 & b & 0.37 & 6.05 & 1311 & 0.71 \\
4 & a & 0.25 & 1.39 & 933.6 & 0.94 \\
5 & a & 0.20 & 2.43 & 1062 & 0.94 \\
6 & a & 0.72 & 6.85 & 1693 & 0.90 \\
\bottomrule
\end{tabular}
\end{center}
\end{table}

\begin{table}
  \caption{Listing of systems that formed by the abduction of a planet
    from another star. Each of these stars was initially without any
    planets, but one or two planets were captured from another system.
    In two cases (\#7 and \#10) two planets were captured.  }
\label{Tab:capturedplanets}
\begin{center}
\begin{tabular}{ll*{4}{>{\raggedleft\arraybackslash}p{1cm}r}}\toprule
 % & \multicolumn{7}{c}{$N_P$} & \\ \cmidrule(r){2-8}
System & Planet & M $(M_{sun})$ & m $(M_{jup})$ & a (au) & e  \\ \midrule
7 & a & 9.16 & 0.28 & 1544 & 0.68 \\
7 & b & 9.16 & 0.50 & 1161 & 0.22 \\
8 & a & 6.70 & 14.69 & 3332 & 0.50 \\
9 & a & 3.61 & 0.95 & 891.5 & 0.77 \\
10 & a & 0.47 & 0.03 & 1100 & 0.66 \\
10 & b & 0.47 & 0.12 & 1208 & 0.71 \\
11 & a & 0.55 & 0.38 & 192.5 & 0.60 \\
\bottomrule
\end{tabular}
\end{center}
\end{table}

The abducted planets in Table\,\ref{Tab:capturedplanets} appear to
have large semimajor axes and a broad range in eccentricities.  Such
abduction explains the observed orbital parameters of the dwarf planet
Sedna in the solar system \citep[see][]{2015MNRAS.453.3157J}.
As an alternative to abduction, a free-floating planet could in principle be
captured by a star or planetary system. Capturing free-floating
planets was also studied in \cite{2018MNRAS.473.1589G}, who argued that
these systems may not be uncommon, but that they would have a wide
range in eccentricities and typically large semimajor axes
\citep{2012ApJ...750...83P}. In our simulations no free-floating
planets were captured, and we do not expect this to be a common
process because 80\% of the ejected planets escape promptly from the
cluster (see \S\,\ref{Sect:FFPs}).

\subsection{Characteristics of colliding planets}\label{Sect:Collisions}

One important aspect of planets is their finite size, which makes them
prone to collisions. A total of 75 planets in our simulations collide
with another planet or with the parent star. In our simulations,
collision with the parent star is not treated realistically in the
sense that we ignored tidal effects. We compensate for this by
adopting a size of 1\,au for planetary-hosting stars. As a result, we
overestimate the number of collisions with the host star and we do not
acquire hot Jupiter planets. We, therefore, focus on the collisions
that occur between planets.

In Table\,\ref{Tab:mergers1} and Table\,\ref{Tab:mergers2} we list the
mergers that occurred in our simulations sorted in the moment of the
collision.  In Table\,\ref{Tab:mergers1} we show the pre-collision
parameters of the two planets that participate in the collision,
whereas in Table\,\ref{Tab:mergers2} we list the orbital parameters of
the merger product.

The orbital parameters for the pre-merger planets are derived from the
last snapshot before the merger occurred, which can be up to 1000
years before the actual event.  The mean mass of the primary in a
colliding planet pair is 1.14\,\MJup\, and a secondary of
0.36\,\MJup. The resulting merger product is 1.5\,\MJup.  During the
calculation, 34 planets collided in a total of 19 events.  Several
planets experienced multiple collisions, causing the planet mass to
increase very effectively and causing the planet to migrate closer toward
the host star.  These multiple mergers all tend to occur in relatively
short succession.

Most mergers tend to occur between neighboring planets, but there are
seven occasions where one or more intermittent planets are skipped. In
particular the event at $t=5.91$\,Myr is interesting because in this
case, the outermost planet collides with the innermost planet.
Although not taken into account in our calculations, such close
encounters among planets could lead to the capture of one planet by
the other, giving rise to a binary planet as was observed in Kepler
1625 \citep{2018arXiv181011060H}.

\begin{table*}
\caption{Orbital elements of the merging planets. For each merger the
  time of the snapshot saved just before the merger is given. For
  every planet the index of the planetary system is given with a
  letter denoting the position of the planet in the system (from the
  innermost planet 'a' to the outermost planet 'f').  We define
  the inclination of a planet with respect to the initial orbital
  plane of the closest planet to the star.}
\label{Tab:mergers1}
\begin{center}
\begin{tabular}{>{\raggedright\arraybackslash}p{1cm}>{\raggedright\arraybackslash}p{0.4cm}*{4}{>{\raggedleft\arraybackslash}p{1cm}}>{\raggedright\arraybackslash}p{0.4cm}*{4}{>{\raggedleft\arraybackslash}p{1cm}}}\toprule
 % & \multicolumn{7}{c}{$N_P$} & \\ \cmidrule(r){2-8}
  Time (Myr) & id & M ($M_{jup}$) & a (au) & e & i ($^\circ$)& id & M ($M_{jup}$) & a (au) & e & i ($^\circ$)\\
  \cmidrule(r){2-6} \cmidrule(r){7-11}
3.10 & 1e & 0.30 & 213.7 & 0.68 & 14.4 & 1d & 0.15 & 105.5 & 0.18 & 125.0  \\
3.14 & 1e & 0.46 & 78.3 & 0.70 & 31.8 & 1c & 0.08 & 33.4 & 0.38 & 100.2  \\
3.28 & 1e & 0.54 & 37.4 & 0.39 & 155.2 & 1b & 0.05 & 13.5 & 0.29 & 87.5  \\
3.85 & 2e & 1.30 & 150.5 & 0.31 & 32.4 & 2d & 0.62 & 118.2 & 0.21 & 37.2  \\
4.24 & 2e & 1.93 & 129.0 & 0.18 & -31.1 & 2b & 0.19 & 92.1 & 0.36 & 37.3  \\
4.96 & 3e & 1.03 & 234.0 & 0.50 & 4.0 & 3d & 0.51 & 90.9 & 0.22 & 7.3  \\
5.13 & 4f & 0.76 & 401.5 & 0.51 & -38.3 & 4c & 0.13 & 130.5 & 0.35 & -6.1  \\
5.31 & 5d & 0.58 & 130.3 & 0.66 & -4.5 & 5c & 0.25 & 115.6 & 0.20 & -9.4  \\
5.36 & 6e & 0.49 & 160.6 & 0.57 & 7.3 & 6d & 0.24 & 87.7 & 0.29 & 18.9  \\
5.49 & 7e & 1.12 & 222.3 & 0.69 & 38.7 & 7d & 0.56 & 161.0 & 0.54 & -5.0  \\
5.83 & 8d & 3.93 & 295.4 & 0.54 & 0.0 & 8a & 0.39 & 96.8 & 0.99 & 7.1  \\
5.85 & 8d & 4.31 & 255.7 & 0.56 & 6.1 & 8c & 1.61 & 296.1 & 0.52 & 5.2  \\
5.91 & 9e & 0.37 & 223.4 & 0.38 & 0.0 & 9a & 0.04 & 215.0 & 0.84 & -10.2  \\
5.98 & 10e & 0.30 & 101.7 & 0.39 & 18.2 & 10d & 0.17 & 133.3 & 0.42 & -10.5  \\
6.09 & 5d & 0.82 & 121.9 & 0.57 & -26.1 & 5b & 0.12 & 44.8 & 0.63 & 5.9  \\
6.20 & 11e & 0.81 & 111.1 & 0.62 & -12.6 & 11d & 0.39 & 44.4 & 0.44 & -13.9  \\
7.55 & 12d & 0.13 & 128.1 & 0.19 & -8.9 & 12b & 0.04 & 98.4 & 0.60 & 5.1  \\
7.71 & 13f & 0.17 & 214.9 & 0.61 & 46.2 & 13e & 0.09 & 42.6 & 0.92 & 24.9  \\
8.25 & 14f & 2.29 & 301.0 & 0.15 & 1.1 & 14e & 1.20 & 127.8 & 0.01 & -1.9  \\
\bottomrule
\end{tabular}
\end{center}
\end{table*}

\begin{table}
\caption{Orbital elements of the planets resulting from a merger.}
\label{Tab:mergers2}
\begin{center}
\begin{tabular}{l*{6}{>{\raggedleft\arraybackslash}p{0.75cm}}}\toprule
 % & \multicolumn{7}{c}{$N_P$} & \\ \cmidrule(r){2-8}
Time (Myr)& id a & id b & M $(M_{jup})$ & a (au) & e & i ($^\circ$)\\ \midrule
3.10 & 1e & 1d & 0.46 & 78.8  & 0.70 & 100.5  \\
3.14 & 1e & 1c & 0.54 & 37.2  & 0.40 & 93.6  \\
3.28 & 1e & 1b & 0.59 & 19.0  & 0.14 & 92.7  \\
3.85 & 2e & 2d & 1.93 & 129.9 & 0.23 & 34.6  \\
4.24 & 2e & 2b & 2.12 & 102.9 & 0.08 & 32.6  \\
4.96 & 3e & 3d & 1.54 & 137.5 & 0.21 & 5.7  \\
5.13 & 4f & 4c & 0.89 & 256.2 & 0.26 & -9.5  \\
5.31 & 5d & 5c & 0.82 & 105.1 & 0.55 & -7.3  \\
5.36 & 6e & 6d & 0.73 & 105.7 & 0.27 & 13.0  \\
5.49 & 7e & 7d & 1.69 & 172.4 & 0.64 & 14.4  \\
5.83 & 8d & 8a & 4.31 & 236.3 & 0.40 & 5.4  \\
5.85 & 8d & 8c & 5.92 & 220.9 & 0.50 & 5.4  \\
5.91 & 9e & 9a & 0.41 & 166.6 & 0.24 & -10.1  \\
5.98 & 10e&10d & 0.46 & 98.7  & 0.31 & 1.4  \\
6.10 & 5d & 5b & 0.94 & 94.8  & 0.54 & -8.0  \\
6.20 & 11e&11d & 1.21 & 63.2  & 0.48 & -5.4  \\
7.55 & 12d&12b & 0.16 & 107.4 & 0.06 & -6.3  \\
7.71 & 13f&13e & 0.26 & 73.0  & 0.14 & 27.0  \\
8.25 & 14f&14e & 3.50 & 255.7 & 0.10 & -1.3  \\
\bottomrule
\end{tabular}
\end{center}
\end{table}

\subsection{Production of free-floating planets}\label{Sect:FFPs}

By the time the cluster has reached an age of 11\,Myr the total mass
in bound planets was reduced from $\sim 3527$\,$M_{\rm jup}$ to $\sim
2915$\,$M_{\rm jup}$ (see Table\,\ref{Tab:SPZ2016}). Planets have been
lost by their parent star via encounters with other stars (see
\S\,\ref{Sect:FFPs}) , internal planet-scattering ($\sim 60$), by the
mass loss of their host stars, and through collisions with the star
(75; see \S\,\ref{Sect:Collisions}) of collided with another planet
(14).  Once liberated, free floaters may remain bound to the cluster
(75 planets) or escape its gravitational potential (282, see
Table\,\ref{Tab:SPZ2016}).  In total 357 planets (out of 2522) were
liberated from the gravitational pull of their parent star.  In
\S\,\ref{Sect:Captures} we discussed the possibility of captured
planets, but this was not the fate of any of the free-floating
planets, because all captured planets were exchanged during a close
encounter and always bound to at least one star.

\begin{figure}[!htb]
  \centering
  \scalebox{1.0}{
%Added by TeX Support
    \input{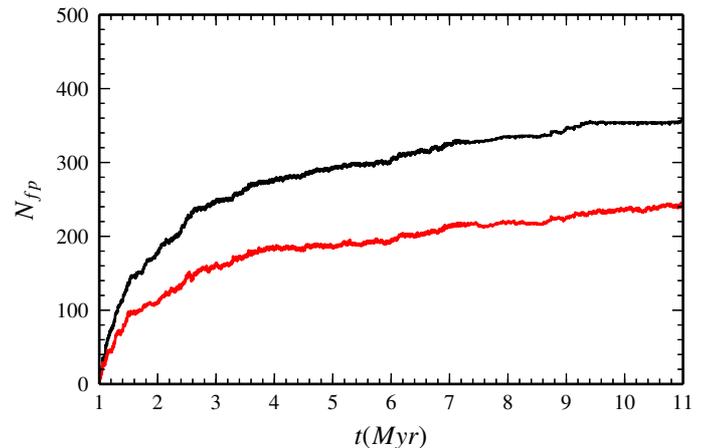}
  }
\caption{Number of free-floating planets $(N_{fp})$ as a function
  of time.  The solid curve (black) indicates all free planets; the red
  curve indicates the subset of free floaters that also escape the
  cluster.  }
\label{fig:nfp_vs_t}
\end{figure}

In Fig.\,\ref{fig:nfp_vs_t} we present the number of free-floating
planets as a function of time.  The majority of free floaters (67\%)
leave the cluster within a crossing time ($\sim 1$\,Myr) after being
liberated from their host star. The other $\sim 33$\% remain bound to
the cluster for an extended period of time and leave the cluster on a
much longer timescale, at a typical escape rate of $\sim 8$ planets
per Myr.

% KS TEST:
%bound/unbound ff: (0.14141414141414144, 0.25517316600868384)
%bound/all ff: (0.16161616161616166, 0.13619189865976927)
%unbound/all ff: (0.1111111111111111, 0.54934699150809463)I

\begin{figure}[!htb]
  \centering
  \scalebox{1.0}{
%Added by TeX Support  
    \input{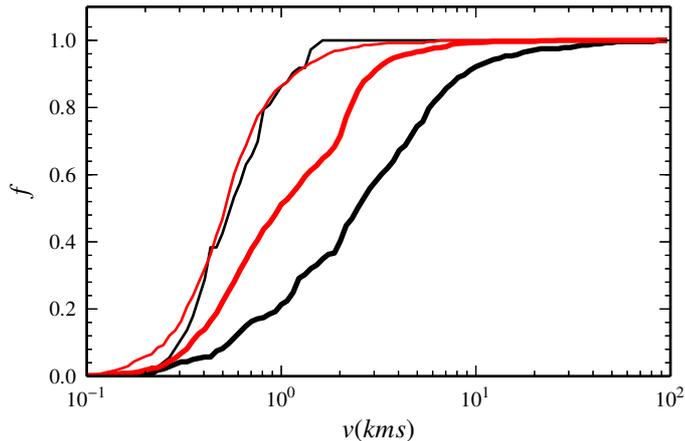}
    }
\caption{Cumulative distribution (normalized) of the velocity of
  planets (black curves) and stars (red curves) at an age of
  11\,Myr. Planets and stars bound to the cluster are plotted with a
  thin line; the thick curves indicate the unbound objects.}
    \label{fig:star_planet_vel}
\end{figure}

In Fig.\,\ref{fig:star_planet_vel} we present the cumulative
distributions of the velocity of bound and unbound stars and planets;
for the planets we make a distinction between free-floating planets
that remain bound to the cluster and those that escape.  The
distributions for the stars and planets at an age of 11\,Myr that are
still bound to the cluster show only slight differences (thin
lines). Both velocity distributions are statistically
indistinguishable (with a KS-statistic of $0.23$). The population of
unbound planets, however, tend to have much higher velocities (of
$\sim 3^{+5.6}_{-1.2}$\,km/s) than the stars ($\sim
1^{+1.3}_{-0.6}$\,km/s).  This is not unexpected because planets tend
to be launched from the stars with their orbital speed, which gives
rise to higher mean escape velocity, whereas most stars escape by
dynamical evaporation
\citep{1995MNRAS.276..206F,1999CeMDA..73..179P}. This relatively high
space motion of the rogue planets is also reflected in the large
percentage of liberated planets that escape the cluster.

\begin{figure}[!htb]
\centering
\includegraphics[width=1.1\columnwidth]{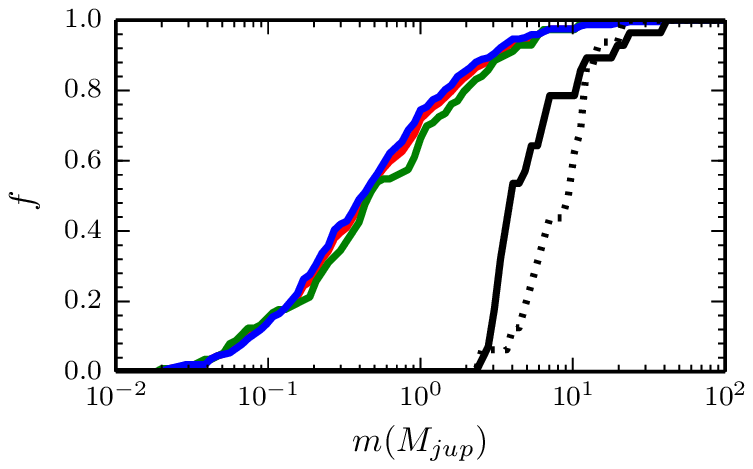}
\caption{Cumulative distribution of the masses of all planets (red),
  the free-floating planets that are bound to the cluster (green), and
  those unbound from the cluster (blue).  These three curves are
  statistically indistinguishable.  The dotted curve indicates the mass
  distribution of 16 observed potential free-floating planets from
  \cite{2005ApJ...635L..93L,2010ApJ...709L.158M,2000Sci...290..103Z,2012A&A...548A..26D,2013ApJ...777L..20L,
    2014ApJ...783..121G,2014AJ....147...34S,2014ApJ...786L..18L,2014ApJ...785L..14G,2016ApJ...833...96L,2014ApJ...792L..17G,2015ApJ...808L..20G,2016ApJ...821L..15K,2016ApJ...822L...1S}.
  For a different comparison, we introduce a lower mass cutoff to
  the initial sample of planets of $2$\,M$_{\rm Jup}$ and compare this
  with the observed sample (black).  
  }
\label{fig:free_planet_mass_function}
\end{figure}

In Fig.\,\ref{fig:free_planet_mass_function} we present the mass
distribution of free-floating planets. Those that remain bound to
the cluster have statistically the same mass function as those that
escape (KS-statistics of $0.14$) and as the global initial planet
mass function (KS $= 0.11$; see also
Fig.\,\ref{fig:initial_meanmass}). Signifying what we already
discussed in relation to Fig.\,\ref{fig:initial_meanmass} and
Fig.\,\ref{fig:initial_planet_mass_hist}: the ejection of planets is
independent of their mass \citep[see
  also][]{2011MNRAS.411..859M,2012MNRAS.425..680V}.

The mass distributions of free-floating planets in the simulation
differ considerably from the observed mass distribution.
Observational selection effects probably play an important role here
because low-mass free-floating planets tend to be very hard to
discover. We, therefore, introduce a lower limit of $2.5\,M_{\rm Jup}$
to the mass distribution the simulated distribution of free floaters
becomes statistically indistinguishable from the observed sample
(KS-statistic is 0.06).

\begin{figure}[!htb]
  \centering
    \scalebox{1.0}{
%Added by TeX Support
      \input{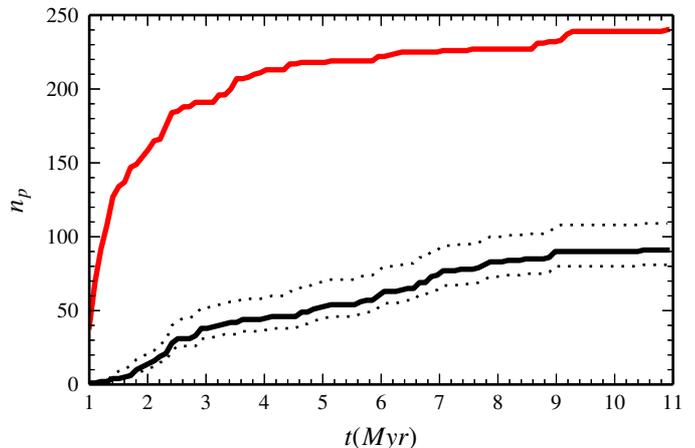}
      }
\caption{Number of planets that became unbound from their host star as
  a function of time. The number of planets that escaped their host
  within 0.5\,Myr following a strong encounter (within 1500\,au, red
  curve) is about twice as large as the planets that escape without
  evidence of having experienced a strong encounter (black curve).
  The dotted black curves indicate the dependency on the timescale
  within which a strong encounter is supposed to lead to ejected
  planets; the lower curves indicate the cumulative distribution for
  planets that are liberated within 1\,Myr of a close encounter,
  whereas the upper curve is for 0.2525\,Myr).  }
\label{fig:planets_time_1_hist}
\end{figure}

In relation to Fig.\,\ref{fig:initial_planet_mass_hist}, we argued
that the lack of a mass-dependency of the production of free-floating
planets is mainly caused by the importance of strong encounters with
other stars rather than internal scattering among planets.  To
quantify this hypothesis we present in
Fig.\,\ref{fig:planets_time_1_hist} the cumulative distributions of
the number of strong and weak encounters.  In this figure strong indicates an
encounter within 1500\,au.  In this analysis, a planet that becomes
free floating within 0.5\,Myr of such a strong encounter is considered
to be liberated as a result of this, otherwise, we consider the planet
to be lost as a result of a weak encounter or the internal
reorganization of the planetary system.

\begin{figure}[!htb]
  \centering
    \scalebox{1.0}{
%Added by TeX Support
      \input{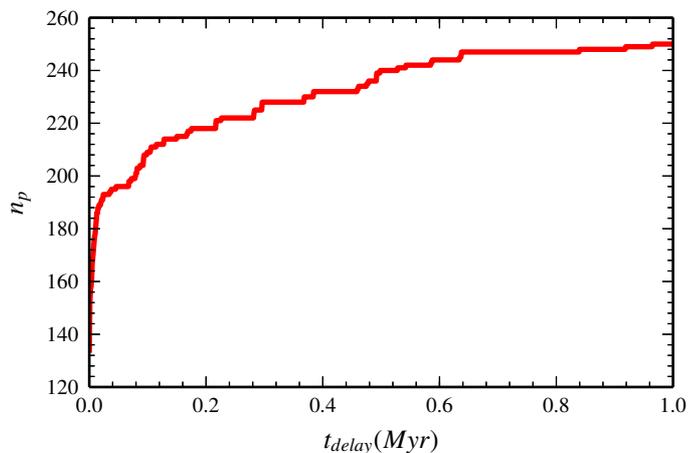}
      }
\caption{Number of planets that escape from their host star as a
  function of the time between a close encounter (within 1500\,au) and
  the moment of escape. The majority of the planets escape promptly
  upon an encounter, but a considerable number require more time,
  up to about a million years.  }
\label{fig:planets_time_2_hist}
\end{figure}

To further understand the importance of strong encounters we present
in Fig.\,\ref{fig:planets_time_2_hist} the delay time distribution of
liberated planets. The majority of those escape promptly upon a strong
encounter with another planetary system or a single star. A
considerable number ($\sim 24$\,\%) require more time (up to about
a million years) before they escape from their host star.  In this
latter population, planetary escape is initiated by the close
encounter, but it requires the planetary system to become dynamically
unstable before the planet is actually ejected. The timescale for
these planetary systems to become unstable appears to be on the order
of a million years.

The number of Jupiter-mass free-floating planets have been estimated to
about 0.25 of the number of main-sequence stars
\citep{2012Natur.481..167C,2017Natur.548..183M}.  This number is
consistent with our findings, even though we adopted that only about
one-third of the stars had planets initially.  If each star would have
a planetary system, our estimates would rise to about $\sim 0.72$\,
free-floating planets per main-sequence star, which would be on the
high side but not inconsistent with the observed estimate of
$1.8^{+1.7}_{-0.8}$ \citep{2011Natur.473..349S}.  Although not taken
into account in this work, the number of free-floating planets produced per
star depends on the moment circumstellar disks start forming planetary
systems, their distribution in mass and orbital parameters, and on the
density and velocity distribution of the young cluster.

\section{Discussion}\label{Sect:Discussion}

We simulated the evolution of a cluster of 1500 stars of which 500 are
orbited by a total of 2522 planets (4, 5, or 6 planets of
0.008\,\MJup\, to 130\,\MJup\, per star in circular planar orbits
between 10\,au and 400\,au). The calculations were performed via the
{\tt Nemesis} script in the Astrophysical Multipurpose Software
Environment \citep{2011ascl.soft07007P,portegies_zwart_simon_2018_1443252,AMUSE}
and include the effects of stellar mass loss and the interactions
between all objects.  We took the initial
conditions from earlier calculations that mimic the mass
and size distributions of the Orion trapezium star cluster
\citep{2016MNRAS.457..313P}.  We stopped the calculations at an age of
11\,Myr, after which we analyzed the population of planets.

In our calculations, we ignored the effect of tidal energy dissipation
between stars and planets. When we started this study we argued that
this effect had minor consequences, but it turned out that 75 of the
planets (3.0\,\%) have a strong interaction with their host star and
34 planets collide with other planets. Tidal interactions are clearly
important and we will improve this in a future version of {\tt
  Nemesis}. Considering these systems as resulting either in a
collision with the parent star or the formation of a hot Jupiter, we
derive a hot-Jupiter formation efficiency of 75 per 500 planetary
systems per 10\,Myr, or 15\% of the planetary systems produce a hot
Jupiter, which is not inconsistent with the rate derived by
\cite{2018arXiv180606601H}.

Our study mainly focuses on the production of free-floating planets.
The planet-ejection probabilities in our simulations are independent
of the mass of the planet, which contradicts earlier
results of \cite{2011MNRAS.411..859M,2014prpl.conf..787D}.  Part of
this result probably depends sensitively on our initial
distributions of planet mass and orbital topology. The choice of
oligarchic growth causes the more massive planets to be further away
from the host star, where they are more vulnerable to perturbations by
passing stars.  This makes the inner planets more prone to being
ejected in the subsequent unstable planetary system that results from
an external perturbation.

Our finding that the probability of escaping the parent star is
independent of planet mass and the birth distance from the star is a
direct consequence of the way in which planets are freed, i.e., in
most cases this is the result of a strong encounter between the
planetary system and another star or planetary system.  In our
simulations, interactions between planets and stars lead to a total of
357 free-floating planets from an initial population of 2522 bound
planets.  This results in 0.24 to 0.70 free-floating planets per
main-sequence star, which is consistent with estimates of the number
of free-floating planets in the Galaxy by \cite{2012Natur.481..167C} and
\cite{2017Natur.548..183M}.

An important reason for the relatively small number of free floaters
is their relatively late formation.  Most interactions occur in the
first 1\,Myr of the evolution of the cluster, and strong dynamical
encounters drive the size evolution of the circumstellar
disks in this phase.  By the time we introduced the planets the stellar density had
already dropped considerably and the number of strong interactions had
subsided.  The absence of planets in the first million years enables them to
survive to a later epoch.  If these disks were already rich in debris
or planets they would have been much more vulnerable to external
perturbations. The mutual interactions between stars in the earliest
cluster evolution $\aplt 1$\,Myr would have been sufficient for
ionizing most planetary systems, leading to a larger population of
free-floating objects.  Such a {\em sola lapis} has recently been
found traversing the solar system \citep{2018MNRAS.479L..17P}.

The distribution of the masses of free-floating planets in our
simulation is indistinguishable from the mass distribution of planets
bound to their host star. This may have interesting consequences for
observations. This comparison may also be made for observed planets.
Our cluster is not old enough to produce free-floating planets by the
copious stellar mass loss in the post-asymptotic giant branch phase,
and it is not a priori clear what effect this would have on the
distribution and ejection of multi-planet systems. But to first order
we argue that the distribution of free-floating planets is the same as
that of bound planets.

\section{Conclusions}\label{Sect:Conclusions}

We simulated the evolution of the Orion Trapezium star cluster
including planets.  The calculations start with initial conditions
taken from earlier calculations at an age of 1\,Myr from
\cite{2016MNRAS.457..313P} by converting circumstellar disks into
planetary systems and were continued to an age of 11\,Myr.  Our
calculations, performed with AMUSE, include the effects of stellar mass loss, collisions, and
the dynamics of the stars and planets. The orbits of the planets are
integrated using a symplectic direct $N$-body code whereas the stellar
dynamics is resolved using a direct Hermite $N$-body code.

Realizing that we study a chaotic system based on the result of only
two simulations, one without stellar interactions and one that
included interactions between the planets and the stars, we
nevertheless feel sufficiently bolstered by our results to report a
number of conclusions.  Each of these conclusions
is based on the results obtained from the simulation in which all
interactions between stars and planets are taken into account.  The
results enumerated below are therefore rather empirical, although, as
argued in the main text, some of these conclusions may be fundamental.
All conclusions, however, are a result of the complicated interplay
between initial conditions and simulations, and it is sometimes hard
to disentangle the two.

\begin{itemize}
\item[]{\bf Conclusions regarding planet stability}
  \begin{itemize}
    \item[$\bullet$] The majority of planets ($\sim 70$\%) experience
      a change in their orbits (in eccentricity or semimajor axis) of
      less than 5\%.
    \item[$\bullet$] A small number of $\sim 10$\,\% planets acquire
      a high ($\apgt 0.8$) eccentricity.  This is not necessarily
      caused by stars passing closely, but in the majority of cases
      repeated small perturbations within the cluster and subsequent
      secular evolution within the planetary system drives these high
      eccentricities.
    \item[$\bullet$] High eccentricities are also induced by
      collisions between planets and in the orbits of captured planets.
    \item[$\bullet$] The innermost planets (at 10\,au) experience a
      comparable relative variation in their final orbital parameters
      (in particular the eccentricity and inclination) due to
      encounters, perturbations, and internal secular evolution as
      wider systems.
    \item[$\bullet$] The probability for a planet to escape is
      independent of its mass or semimajor axis.  Low-mass planets
      that are born relatively close to the parent star are only
      marginally more prone to ejection than more massive planets born
      further out \citep[see also,
      ][]{2011MNRAS.411..859M,2012MNRAS.425..680V}.  This result,
      however, probably depends sensitively on the initial
      orbital distributions and masses of the planets. Comparing
      observed planet-mass distributions and those that survived in a
      planetary systems may then provide interesting constraints on
      the initial planet mass function.
    \item[$\bullet$] Seventy-five planets (3.0\%) collide with their host
      star. This number, however, strongly depends on our adopted
      stellar collision radius and will change when tidal evolution is
      properly taken into account, but we still expect that collisions
      between a planet and its host star are rather frequent.
      Although our collisions are not taken into account
      realistically because of the large stellar size we adopted, these
      systems would be eligible to the formation of hot Jupiter planets
      at a rate of $\sim 0.015$ per star per Myr.
    \item[$\bullet$] The widest planetary systems in our simulations
      tend to be formed either by ejecting planets on very wide and
      highly eccentric orbits or by capturing a planet from another
      star. Both methods seem to be equally important, but the
      captured planets tend to have somewhat lower eccentricity.
  \end{itemize}
\item[]{\bf Conclusions regarding planetary escapers}
  \begin{itemize}
    \item[$\bullet$] A total of 357 planets (out of 2522 or $\sim
      16.5$\,\%) become unbound from their parent star.
    \item[$\bullet$] Out of 357, 282 ($\sim 80$\%) of the free floating
      planets promptly escape the cluster upon being unbound from
      their parent stars.
    \item[$\bullet$] The probability for a planet to escape is
      independent of its mass.  As a consequence, the mass function of
      free-floating planets and the mass function of bound planets are
      indistinguishable from the initial distribution of planet
      masses.
    \item[$\bullet$] At the end of our simulations systems with 3
      planets were rare compared to systems with 1 or 2 planets,
      or systems with 4 or more planets.  Once a star loses planets,
      it tends to lose 3 or more 
      \citep[consistent with Table 9 of][]{2017MNRAS.470.4337C}.
  \end{itemize}
\item[]{\bf Conclusions regarding planet collisions}
  \begin{itemize}
    \item[$\bullet$] Thirty-four planets (1.3\%) experienced a collision with
      another planet.
    \item[$\bullet$] The collision probability between two planets is
      independent of planet mass.
    \item[$\bullet$] The orbits of planet-planet collisions have a
      mean eccentricity of $0.33\pm 0.19$ and a relative inclination
      of $20^\circ\pm 35^\circ$.
    \item[$\bullet$] Instead of colliding, some of those events may
      lead to the tidal capture of one planet by another. This would
      lead to the formation of a binary planet, or moon, as was
      observed in Kepler 1625B \cite{2018AJ....155...36T}.
    \item[$\bullet$] It is generally the outermost planet that
      collides with a planet closer to the parent star. This inner
      planet is not necessarily the next nearest planet.
    \item[$\bullet$] Planets regularly engage in a cascade of
      collisions. These chain-collisions are initiated by a dynamical
      encounter with another star.
  \end{itemize}
\item[]{\bf Conclusions regarding the host star clusters}
  \begin{itemize}
    \item[$\bullet$] The host star ejects 240 (67\% of the ejected planets, 10\% of all
      planets) planets with a delay
      of $0.1$--$0.5$\,Myr after the last strong encounter with
      another cluster member.
    \item[$\bullet$] Young $\sim 10$\,Myr old star clusters harbor a
      rich population of free-floating planets.  About one-third of the
      free-floating planets remain in the cluster for more than a
      dynamical timescale, up to the end of the simulation.  The
      number of free floaters in these clusters can be as high as $40$
      planets for stars between 0.9\,\MSun\, and 1.1\,\MSun, or
      $25$\,\% of the main-sequence stars \citep[consistent with
        estimates by][]{2012Natur.481..167C}.
  \end{itemize}
\end{itemize}

A large number (30\%) of planetary systems are affected by the
presence of the other stars in the cluster, but only $\sim 10$\% of
those will leave a recognizable trace that allows us to reconstruct
the dynamical history based on the topology of the inner planets.  For
the majority of planetary systems observed today, current instruments
are unable to discern the dynamical history because we only observe
the inner most planets, rather than the outer parts where dynamical
effects are most pronounced. It would require observation of a
exo-Kuiper belt to be able to establish the past dynamical history of
the planetary system.  Possibly the easiest way to perceive the
dynamical history of a planetary system is preserved in collision
products between planets.  We argue that in more than 3\% of the
planetary systems collisions between planets are initiated by external
dynamical perturbations.  From the $\sim 4000$ planetary systems known
today we then expect more than one hundred to host a collision
product.

About 16\% of planets eventually become dissociated from their parent
star due to interactions with other cluster members or internal
reorganization of the planetary system. These ejected planets become
free floaters. The majority of those ($\apgt 80$\%) leave the cluster
within a crossing time scale, the rest lingers around the cluster
potential and are subject to a slower evaporation process driven by
mass segregation. We therefor expect star clusters to be relatively
poor in free floating planets. The Galactic field, on the other hand,
is contains about 1/4-th of the number of free floating planets as
there are main-sequence stars. The Galaxy is then composed of some
$5\times 10^{10}$ free floating planets, of which only a dozen are
observed.

%% No planets in our simulations were hurt or maimed in this study.

%\FloatBarrier
\section*{Acknowledgment}

We are grateful to the anonymous referee for many interesting comments
that helped us improve the manuscript.  In this work we used the
following packages: AMUSE
\citep{2011ascl.soft07007P,portegies_zwart_simon_2018_1443252} (which
is also available at {\tt http://amusecode.org}), Hermite0
\citep{2014DDA....4530301M}, Hop \citep{2011ascl.soft02019E}, Huayno
\citep{2012NewA...17..711P}, matplotlib \citep{2007CSE.....9...90H},
numpy \citep{Oliphant2006ANumPy}, Python \citep{vanRossum:1995:EEP},
and SeBa \citep{2012ascl.soft01003P}.
The calculations ware performed using the LGM-II (NWO grant \#
621.016.701) and the Dutch National Supercomputer at SURFSara (grant
\# 15520).

%\bibliographystyle{plain}
%\bibliographystyle{aa}
%\bibliography{references}

\appendix

\section{Validation}\label{Sect:AppValidation}
We analyze the accuracy of the hybrid {\tt Nemesis} strategy as
a function of the interaction time step $dt_{\tt Nemesis}$.

\subsubsection{Determining the optimal {\tt Nemesis} time step}\label{Sect:dtNemesis}

This {\tt Nemesis} time step ($dt_{\tt Nemesis}$) numerically
associates two important factors: how often forces between subsystems
are calculated and a measure for the interaction distance
between individual particles ($d_{\tt Nemesis}$). If two particles are
separated by less than this interaction distance ($d_{\tt Nemesis}$) a
new subsystem is created within which the interaction between
particles is resolved with a separate $N$-body integrator. In
principle we create a new subsystem with its own individual $N$-body
solver. In practice, however, many of these individual subsystem
$N$-body solvers are the same code.

If one particle is spatially separated from several other particles in
a subsystem by a distance larger than the interaction distance,
$d_{\tt Nemesis}$, this particle is removed from the subsystem and
incorporated in the global cluster integration code.  For the physics
it makes no difference if a particle is part of the global system or
of a subsystem.  However, the integrator used for any of the
subsystems is symplectic and generally more accurate by adopting higher
order and a smaller time step, whereas the global $N$-body code adopts
larger time steps and is not symplectic.

There is no specific requirement for any particle to be integrated
either by the integrator of a subsystem or by the global integrator. The
choice of the domain to which the particle belongs is purely based on
geometry and the adopted demands for accuracy and precision.  In
practice, the entire cluster including all the planets could either be
integrated by the single global fourth order Hermite code or by one of
the symplectic $N$-body codes of the subsystem. The choice of which particle
is integrated by what integrator is then only decided on terms of
accuracy, precision, and performance.

As a general note, however, the global $N$-body code tends to be less
accurate because of larger time stepping and non-symplectic, whereas the
subsystem codes adopt rather small shared time steps with a symplectic
integrator.  As a consequence, we prefer to keep particles that belong
to a single planetary system in the same integrator.

\begin{figure}[!htb]
\centering
\includegraphics[width=0.53\textwidth]{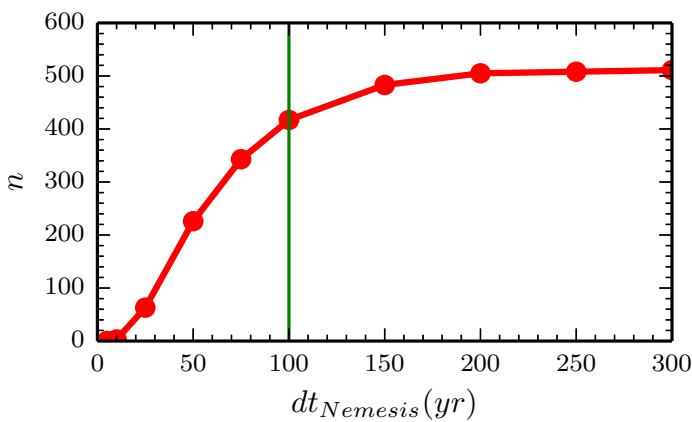}
\caption{Number of the initial intact planetary systems as a function of
  the {\tt Nemesis} time step. The chosen time step of 100\,yr is shown
  as a green vertical line. The time step is not optimal for this
  criterion, but was chosen as it gives better accuracy and a higher
  computational speed.}
\label{fig:validation_dt2}
\end{figure}

The number of stars and planets that are embedded within a single
subsystem depends on $dt_{\tt Nemesis}$ (and therefore on $d_{\tt
  Nemesis}$).  In Fig.\,\ref{fig:validation_dt2} we show how this
number varies as a function of $dt_{\tt Nemesis}$.  For very small
values of $dt_{\tt Nemesis}$, all the stars and planets are integrated
by the global $N$-body integrator, and the number of subsystems $n$
drops to 1, in the extreme. On the other hand, if $dt_{\tt Nemesis}
\apgt 200$\,yr all initial planetary systems are recognized as
individual subsystems and assigned their own integrator. In that case, the number of subsystems grows to the actual number of planetary
systems we initialized plus one for the global $N$-body system, and $n$
approaches to a value of $501$.  We draw a vertical line at $dt_{\tt
  Nemesis} = 100$\,yr, which corresponds to our adopted Nemesis
time step. For this value, a total of about 400 $N$-body integrators are
being initialized and run concurrently.

\subsubsection{Subsystem size criterion in Nemesis}

The analysis performed in the previous section is calculated on a
static initial realization without evolving the cluster
dynamically. In \S\,\ref{Sect:dtNemesis} we demonstrated that at a larger
time step individual planetary systems are consistently captured in a
subsystem. A larger time step is also preferred because this requires
fewer interaction steps between the subsystems and the other
particles. The evolution of the cluster, however, is dynamic and as a
consequence, the value of $dt_{\tt Nemesis}$ should be dynamics to
warrant the accuracy and efficiency of the {\tt Nemesis} method.  We
tested this hypothesis by integrating the cluster for $0.1$\,Myr with
various values of $dt_{\tt Nemesis}$.  After this time we measured the
radius of the largest resolved subsystem.  These largest resolved
subsystems tend to slow down the integration because they are likely to
be composed of a larger number of particles (stars and planets).  Such
large subsystems may cause the entire calculation to wait for the
integration of the large subsystem.  the calculation becomes
progressively slower when more particles are incorporated in the
subsystem.  Eventually, this may continue until all the particles are
embedded in a single subsystem, which is beyond the purpose of the
{\tt Nemesis} module.

\begin{figure}[!htb]
\centering
\includegraphics[width=0.53\textwidth]{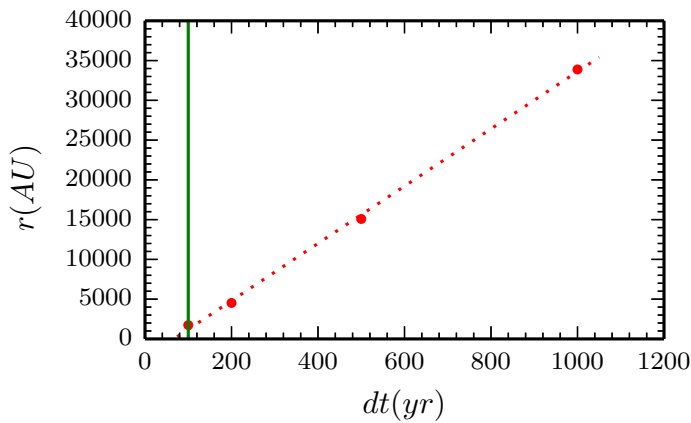}
\caption{Size of the largest subsystem as a function of $dt_{\tt
    Nemesis}$.  after 0.1\,Myr of evolution.  The vertical green line
  indicates the adopted value of 100\,yr.}
\label{fig:validation_dt3}
\end{figure}

In Fig.\,\ref{fig:validation_dt3} we present the measured size of
subsystems as a function of $dt_{\tt Nemesis}$.  The optimum is
reached for a $dt_{\tt Nemesis} \simeq 100$\,yr, which results in a
maximum radius for subsystems of $\sim 1738$\,au. The choice of a time
of $dt_{\tt Nemesis} \simeq 100$\,yr results in the most efficient
calculation of the entire stellar system while at the same time it
results in the lowest energy error.  With this time step our
calculations conserve energy better than one part in $10^{4}$ per
planetary system per million years, which is sufficient to preserve the phase
space characteristics of $N$-body systems for the 10\,Myr over which
we performed the simulation \citep{2014ApJ...785L...3P}.

The two criteria, i.e., (1) keep each initial planetary system in a single
subsystem and (2) prevent subsystems from boundless growth, suggest
opposing optimal values for the Nemesis time step $dt_{\tt Nemesis}$.
Both criteria appear to match for $dt_{\tt Nemesis} \simeq 100$\,yr,
which is the value we adopt for all further calculations.

\subsubsection{Validation of Nemesis on individual planetary systems}

Apart from tuning the performance and accuracy of the compound {\tt
  Nemesis} integrator, we also validated this code in a more practical
application. For this we opted to study the evolution of a system
of five\ planets that is orbited by another second star of $1$M$_{\odot}$
with a semimajor axis of $1500$\,au, an eccentricity of $0.5,$ and
an inclination of $90\degree$.  The planetary system is generated
using the oligarchic growth model for a 1\,\MSun\, star with a 400\,au
disk of 0.1\,\MSun. The simulations were performed via two distinct
methods: (1) using {\tt Nemesis} and (2) integrating all objects in a
single $N$-body code. The {\tt Nemesis} method requires two codes: one
for the planetary system and one for the center of mass of the
planetary system and the orbiting secondary star.

\begin{figure}[!htb]
  \centering
  \scalebox{1.0}{
%Added by TeX Support  
    \input{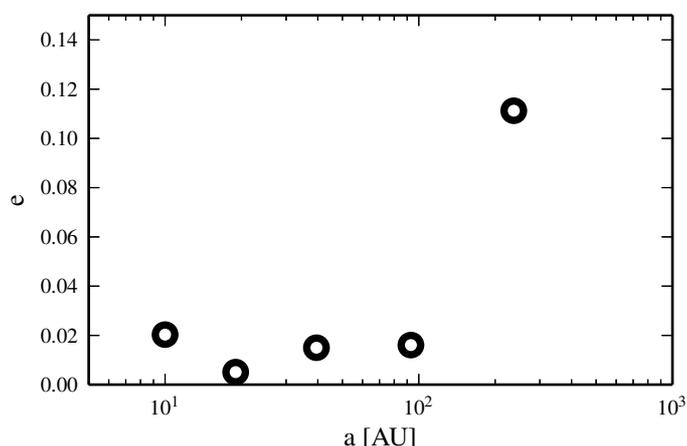}
    }
\caption{Eccentricity as a function of semimajor axis for a planetary
  system orbited by a secondary star of 1\,\MSun\, after $0.5\,Myr$ of
  integration using {\tt Nemesis} (big black bullets) and the single
  8th order symplectic integrator in {\tt Huayno} (smaller white
  bullets). The final eccentricity of the planets in the direct
  integration and the component method are indistinguishable in the
  figure, with an absolute mean error $<2\times10^{-4}$ for each of
  the planets.  }
\label{fig:validation_ae}
\end{figure}

For both integrators, we selected the eighth order symplectic
integrator in {\tt Huayno}.  The two codes
communicate using a Nemesis time step of $dt_{\tt Nemesis} = 100$\,yr.
For comparison, we also integrated these planetary systems with the
same integrator, but all the objects stars and planets are in the same
computational domain.  In Fig.\,\ref{fig:validation_ae} we present the
eccentricities of the planets as a function of the semimajor axis at an
age of 0.5\,Myr.

Based on the integration of these planetary systems and the earlier
tests regarding the migration of planets across integrators, we decided
that a {\tt Nemesis} time step of $dt_{\tt Nemesis} = 100$\,yr gives
satisfactory results in terms of accuracy, precision, and performance.

\subsubsection{Energy errors in the composite model}

\begin{figure}[htb]
  \centering
  \scalebox{1.0}{  
%Added by TeX Support
    \input{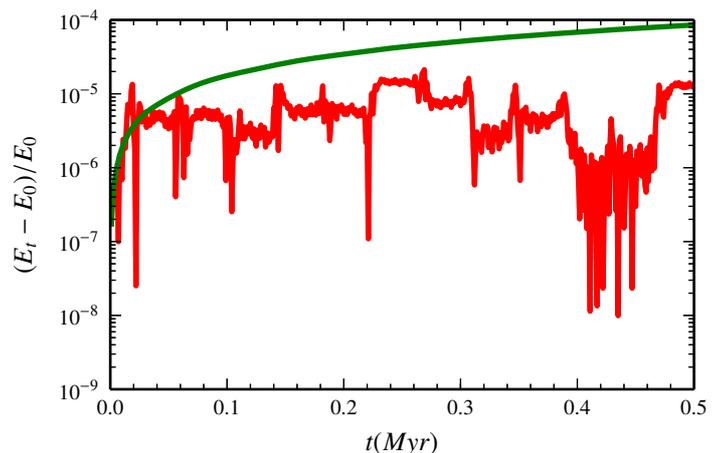}
    }
\caption{Total energy error as a function of time for a validation
  simulation consisting of a single star orbiting a system of 5
  planets.  The energy error of our method (in red) is compared to the
  results obtained using a 4th order Hermite code for all particles
  (green).  The time evolution of the energy error is more erratic in
  the Nemesis method because of the close interactions of the orbiting
  star. The overall error, however, remains rather constant over a
  long timescale, whereas for the Hermite method the energy error is
  smoother but clearly grows with time.}
\label{fig:simplecticity}
\end{figure}

To determine the reliability of the {\tt Nemesis} for planetary system
evolution, we also investigated the evolution of the energy error.  We
performed this test for the same model as in the previous section, using
an isolated planetary system composed of five planets and one perturbing
star in a wide orbit. We simulated this system using our method and a
fourth order Hermite integrator using a time step of $dt_{\tt Nemesis} =
100$\,yr.  The resulting evolution of the energy error is presented
in In Fig.\,\ref{fig:simplecticity}.

In Fig.\,\ref{fig:simplecticity} we show the results of the two
calculations, one with a fourth order Hermite integrator (green), which
is not symplectic. The other calculation (red curve) is performed
using {\tt Nemesis} in which we combine an eighth-order symplectic
integrator for the planetary system with the fourth-order Hermite
integrator for the binary system.  The energy error in the Hermite
(green curve) grows monotonically, which is the typical response for a
non-symplectic integrator, such as the adopted Hermite scheme.  The
evolution of the energy error in the hybrid integrator does not grow
on a secular timescale. The evolution of the energy error is rather
erratic with sharp peaks to low values as well as high values but
stays stable overall. The secular growth of {\tt Nemesis} is much
smaller than the single Hermite integrator. This is mainly caused by
the fact that the largest energy errors are generated while
integrating the planetary system, which, in the Hermite integration
(green curve) drives the energy error.  An additional advantage is
that the calculation with the hybrid {\tt Nemesis} method took about
ten minutes on a workstation, whereas the Hermite scheme (green curve)
took 18 hours.

Based on the results presented in Fig.\,\ref{fig:simplecticity}, we
conclude that in our method the energy error does not grow with time,
but remains constant for the duration of the calculation. The
Hermite part of the integration does show a monotonic increase of the
energy error, but this error remains below the mean error produced in
the subsystem code, which is symplectic. The overall energy error, therefore, appears well behaved, but eventually, in the long run, the
non-symplectic part of the energy error may start to dominate.

%\end{multicols}
\end{document}